\documentclass{aa}  

\usepackage{graphicx}
\usepackage{xcolor}
\usepackage{textcomp}

\usepackage{txfonts}
\usepackage{natbib}
\bibpunct{(}{)}{;}{a}{}{,} 
\DeclareRobustCommand{\gaia}{\textit{Gaia} }
\DeclareRobustCommand{\teff}{$T_{\mathrm{eff}}$}
\DeclareRobustCommand{\logg}{$\log g$}
\DeclareRobustCommand{\mh}{$\mathrm{[M/H]}$}
\DeclareRobustCommand{\vmic}{$v_{\mathrm{mic}}$}

\DeclareRobustCommand{\vsini}{$v\mathrm{sin}i$}
\DeclareRobustCommand{\ispec}{\texttt{iSpec}}
\DeclareRobustCommand{\daospec}{\texttt{DAOSPEC}}
\DeclareRobustCommand{\gala}{\texttt{GALA}}

\begin{document}

   \title{OCCASO V}

   \subtitle{Chemical-abundance trends with Galactocentric distance and age\footnote{Tables 1, A1, A2 and A4 are only available in electronic form at the CDS via anonymous ftp to cdsarc.u-strasbg.fr (130.79.128.5) or via http://cdsweb.u-strasbg.fr/cgi-bin/qcat?J/A+A/.}}


   \author{J. Carbajo-Hijarrubia\inst{1,2,3}
          \and
          L. Casamiquela\inst{1,2,3,4}
          \and
          R. Carrera\inst{5}
          \and
          L. Balaguer-N\'u\~nez\inst{1,2,3}
          \and
          C. Jordi\inst{1,2,3}
          \and 
          F. Anders\inst{1,2,3}
          \and \\
          C. Gallart\inst{6,7}
          \and 
          E. Pancino\inst{8}
          \and 
          A. Drazdauskas\inst{9}
          \and 
          E. Stonkut{\. e}\inst{9}
          \and 
          G. Tautvai{\v s}ien{\. e}\inst{9}
          \and 
          J.M. Carrasco\inst{1,2,3}
          \and
          E. Masana\inst{1,2,3}
          \and \\
          T. Cantat-Gaudin\inst{10}
          \and
          S. Blanco-Cuaresma\inst{11}
          }

\institute{Institut d'Estudis Espacials de Catalunya (IEEC), Gran Capit\`a, 2-4, 08034 Barcelona, Spain\\
        \email{juan636@fqa.ub.edu}
         \and
        Institut de Ci\`encies del Cosmos (ICCUB), Universitat de Barcelona (UB), Martí i Franquès, 1, 08028 Barcelona, Spain
         \and
        Departament de Física Quàntica i Astrofísica (FQA), Universitat de Barcelona (UB),  Martí i Franquès, 1, 08028 Barcelona, Spain
         \and
        Observatoire de Paris (Meudon) GEPI Batiment 11, Hipparque 5 Place Jules Janssen, 92195, Meudon, France
         \and
        INAF - Osservatorio di Astrofisica e Scienza dello Spazio di Bologna, via Gobetti 93/3, 40129, Bologna, Italy
        \and
        Instituto de Astrofísica de Canarias, E-38205 La Laguna, Tenerife, Spain
        \and
        Universidad de La Laguna, Avda. Astrofísico Fco. Sánchez, E-38205 La Laguna, Tenerife, Spain
        \and
        INAF – Osservatorio Astrofisico di Arcetri, Largo E. Fermi 5, 50125 Firenze, Italy
         \and
        Institute of Theoretical Physics and Astronomy, Vilnius University, Sauletekio av. 3, 10257 Vilnius, Lithuania
         \and
        Max-Planck-Institut für Astronomie, Königstuhl 17, 69117 Heidelberg, Germany
         \and
        Harvard-Smithsonian Center for Astrophysics, 60 Garden Street, Cambridge, MA 02138, USA\\
           }

   \date{}

\abstract{}{}{}{}{} 
  
  \abstract
   {Open clusters provide valuable information on stellar nucleosynthesis and the chemical evolution of the Galactic disk, as their age and distances can be measured more precisely with photometry than in the case of field stars.}
   {Our aim is to study the chemical distribution of the Galactic disk using open clusters by analyzing the existence of gradients with Galactocentric distance, azimuth, or height from the plane and dependency with age.}
   {We used the high-resolution spectra ($R$>60\,000) of 194 stars belonging to 36 open clusters to determine the atmospheric parameters and chemical abundances with two independent methods: equivalent widths and spectral synthesis. The sample was complemented with 63 clusters with high-resolution spectroscopy from literature.}
   {We measured LTE abundances for 21 elements: $\alpha$ (Mg, Si, Ca, and Ti), odd-Z (Na and Al), Fe-peak (Fe, Sc, V, Cr, Mn, Co, Ni, Cu, and Zn), and neutron-capture (Sr, Y, Zr, Ba, Ce, and Nd). We also provide non-local thermodynamic equilibrium abundances for elements when corrections are available. We find inner disk young clusters enhanced in [Mg/Fe] and [Si/Fe] compared to other clusters of their age. For [Ba/Fe], we report an age trend flattening for older clusters (age$<$2.5\,Ga). The studied elements follow the expected radial gradients as a function of their nucleosynthesis groups, which are significantly steeper for the oldest systems. For the first time, we investigate the existence of an azimuthal gradient, finding some hints of its existence among the old clusters (age$>$2\,Ga).}
   {}


   \maketitle
%
\section{Introduction}

Various processes involved in the formation and evolution of the Galactic disk have left their imprint on the chemo-dynamical properties of the stars that populate it. Several tracers have been used to unveil these processes, such as \mbox{H\,{\sc ii}} regions \citep[e.g.,][]{HII12011,Arellano2020,mendez_delgado2022}, planetary nebulae \citep[e.g.][]{PN1,Pn2}, Cepheids \citep[e.g.][]{cepheids1,cepheids2,minniti2020,daSilva2022}, low-mass stars \citep[e.g.][]{MS1, Anders2017}, massive stars \citep[e.g.][]{Daflon2004,Bragan2019}, and star clusters \citep[e.g.,][]{Janes1979,friel2002}. Open clusters (OCs) have proved particularly useful as some of their properties such as age and distance can be more precisely determined than for other tracers. Moreover, the OC population covers almost the full disk lifetime, allowing us to trace the overall disk history of star formation and nucleosynthesis \citep[e.g.,][]{Friel2013}.

Studies based on OCs have provided valuable information about the chemical distribution of the disk stars, such as a decreasing metallicity with increasing Galactocentric position \citep[e.g.,][]{Janes1979,bragaglia2008,sestito2008,D'Orazi2009,jacobson2011}. However, these initial studies were hampered by the small number of systems studied homogeneously \citep[e.g.,][]{pancino2010,jacobson2011}. Other authors have built larger samples by compiling values in the literature, but they were limited by their heterogeneity \citep[e.g.,][]{Carrera_Pancino2010,Yong2012,donati2015,Heiter2015,netopil2016}. These issues have been overcome in the last years by the \gaia mission data \citep{Gaia2016} and the massive ground-based spectroscopic surveys to complement them. \citet{GC_recioblanco2022} studied a sample of 503 OCs older than 100\,Ma within a Galactocentric radius ($R_{\rm GC}$) of $\sim$12\,kpc based on the third \gaia Data Release \citep[DR3,][]{Gaia2022_vallenari}. Unfortunately, \gaia only provides abundances for a few elements due to its medium spectral resolution, $R\sim$11\,500, and the small wavelength coverage, 845 to 872\,nm \citep{Gaia2016}.

High resolution, $R$>20\,000, ground-based spectroscopic surveys are providing radial velocities and chemical abundances with a better precision than \textit{Gaia}. Currently, only GES \citep[\textit{Gaia}-ESO survey,][]{gilmore2012}, APOGEE \citep[Apache Point Observatory Galactic Evolution Experiment,][]{Majewski2017}, and GALAH \citep[GALactic Archaeology with HERMES,][]{desilva2015} have published data. GES and GALAH are sampling the Southern Hemisphere, while APOGEE is observing both hemispheres using twin telescopes and instruments. 
APOGEE and GALAH have not made observations specifically designed to study open clusters, which means that the stars studied are in different evolutionary states and therefore show different abundances due to stellar evolution. In addition, the number of sampled stars varies greatly between OCs, with many of them having abundances for a single star. Of the 150 OCs observed in APOGEE \citep{myers2022}, only 47 have abundances of at least four stars. In the case of GALAH, from the 75 observed systems \citep{spina2021} only 14 OCs have measurements of at least four stars. 

GES used two different instruments: GIRAFFE with a spectral resolution similar to APOGEE and GALAH, and UVES \citep[Ultraviolet and Visual Echelle Spectrograph,][]{dekker2000} with a spectral resolution of $\sim$47\,000 covering a wavelength range between 480 and 680\,nm. Recently, \citet[hereafter GES23]{Magrini2022} have studied a sample of 62 OCs older than 100\,Ma acquired with GES-UVES. Their work provides abundances for 25 chemical elements, including a number of neutron capture elements such as Y, Zr Mo or Pr. 
However, GES has sampled only the southern hemisphere, while several key OCs, such as NGC\,6791, are only available for northern observers. 

Since 2013,  we are systematically observing stars in OCs in the framework of the Open Clusters Chemical Abundances from Spanish Observatories (OCCASO) project. Its main driver is to study the kinematic and chemical evolution of the Galactic disk by measuring radial velocities and detailed elemental abundances. For this purpose, we study at least four red clump (RC) stars by high-resolution spectroscopy, $R$>60\,000, and covering a large wavelength range, 400 to 900\,nm \citep[][hereafter Paper\,I]{occaso1}. Atmospheric parameters and Fe abundances for 115 stars belonging to 18 OCs were determined by \citet[][hereafter Paper\,II]{occaso2}. \citet[][hereafter Paper\,III]{occaso3} extended this by deriving abundances of five Fe-peak and five $\alpha$ elements for the same sample. Radial velocities for 336 stars belonging to 51 OCs have been presented by \citet[][hereafter Paper\,IV]{occaso4}.

The present paper is the fifth of the series directly based on OCCASO spectra. We publish chemical abundances for 36 OCs observed until December 2020, doubling the number of OCs published in Paper\,III. We provide improved abundances for 21 chemical elements and, for the first time in the project, the odd-z elements Na, and Al, the Fe-peak elements Mn, Cu and Zn, and neutron capture elements Sr, Y, Zr, Ba, Ce, and Nd. 

The paper is organized as follows. The target selection, observations, and data reduction are explained in Sect.~\ref{sec:target_selection}. The spectroscopic determination of atmospheric parameters is detailed in Sect.~\ref{sec:target_selection}. The chemical abundance computation methods, solar abundance scale, and cluster abundances are explained in Sect. 4. Trends in the Galactic disk are investigated in Sect. 5. Finally, general conclusions are discussed in Sect. 6.

\section{Observations and methodology}\label{sec:target_selection}
    \subsection{Observational material}

    \begin{table*}[h!]
    \setlength{\tabcolsep}{1mm}
    \begin{center}
    \caption{Properties of the 36 OCCASO OCs in this work.}
    \begin{tabular}{lcccccccccccccccccccccccccccccccc}
    \hline
    Cluster        & $\alpha_{\rm ICRS}$ & $\delta_{\rm ICRS}$  & Age &  Distance & \emph{X} & \emph{Y} & \emph{Z} & $R_{\rm GC}$ &[Fe/H]&[Mg/Fe] & $N$\\
                    & [deg]    & [deg] &  [Ga] &  [pc] & [pc] & [pc] & [pc] & [pc] & [dex] & [dex]\\

    \hline
    Berkeley 17&80.13&30.574&7.2&3341&-3325&252&-214&11668&-0.24$\pm$0.04&0.11$\pm$0.28&4\\
    FSR 0278&307.761&51.021&2.2&1708&63&1695&202&8448&0.15$\pm$0.05&-0.06$\pm$0.03&5\\
    FSR 0850&86.257&24.74&0.5&2232&-2226&-136&-89&10567&-0.01$\pm$0.02&-0.03$\pm$0.12&3\\
    IC 4756&279.649&5.435&1.3&506&406&299&47&7938&-0.03$\pm$0.01&-0.08$\pm$0.01&7\\
    NGC 188&11.798&85.244&7.1&1698&-851&1319&646&9285&-0.03$\pm$0.07&0.04$\pm$0.06&4\\
    NGC 752&29.223&37.794&1.2&483&-324&303&-191&8669&-0.02$\pm$0.02&-0.08$\pm$0.03&7\\
    NGC 1817&78.139&16.696&1.1&1799&-1742&-189&-405&10084&-0.16$\pm$0.03&-0.01$\pm$0.02&5\\
    NGC 1907&82.033&35.33&0.6&1618&-1605&207&8&9947&-0.05$\pm$0.02&0.00$\pm$0.03&4\\
    NGC 2099&88.074&32.545&0.4&1432&-1429&58&77&9769&0.06$\pm$0.03&-0.07$\pm$0.01&8\\
    NGC 2354&108.503&-25.724&1.4&1370&-713&-1158&-163&9127&-0.03$\pm$0.04&-0.03$\pm$0.03&6\\
    NGC 2355&109.247&13.772&1.0&1941&-1744&-753&397&10112&-0.10$\pm$0.04&-0.02$\pm$0.02&6\\
    NGC 2420&114.602&21.575&1.7&2587&-2316&-757&869&10683&-0.22$\pm$0.03&0.02$\pm$0.02&7\\
    NGC 2539&122.658&-12.834&0.7&1228&-713&-971&236&9105&0.05$\pm$0.06&-0.10$\pm$0.03&5\\
    NGC 2632&130.054&19.621&0.7&183&-139&-67&98&8479&0.23$\pm$0.04&-0.09$\pm$0.02&4\\
    NGC 2682&132.846&11.814&4.3&889&-613&-440&470&8964&0.04$\pm$0.04&-0.03$\pm$0.02&8\\
    NGC 6633&276.845&6.615&0.7&424&339&247&61&8004&-0.02$\pm$0.03&-0.06$\pm$0.01&4\\
    NGC 6645&278.158&-16.918&0.5&1810&1739&490&-113&6618&0.12$\pm$0.03&-0.04$\pm$0.05&6\\
    NGC 6705&282.766&-6.272&0.3&2203&1955&1009&-106&6464&0.11$\pm$0.07&-0.03$\pm$0.03&12\\
    NGC 6728&284.715&-8.953&0.6&1829&1638&791&-181&6747&0.02$\pm$0.01&-0.06$\pm$0.03&5\\
    NGC 6791&290.221&37.778&6.3&4231&1423&3903&800&7942&0.15$\pm$0.14&-0.06$\pm$0.25&6\\
    NGC 6811&294.34&46.378&1.1&1161&212&1116&241&8203&-0.03$\pm$0.02&-0.06$\pm$0.02&6\\
    NGC 6819&295.327&40.19&2.2&2765&754&2628&407&8027&0.04$\pm$0.06&-0.05$\pm$0.05&4\\
    NGC 6939&307.917&60.653&1.7&1815&-182&1764&386&8703&0.03$\pm$0.06&0.00$\pm$0.06&5\\
    NGC 6940&308.626&28.278&1.3&1101&376&1026&-137&8029&0.14$\pm$0.05&-0.06$\pm$0.03&6\\
    NGC 6991&313.621&47.4&1.5&577&26&576&15&8333&-0.03$\pm$0.01&-0.06$\pm$0.01&3\\
    NGC 6997&314.128&44.64&0.6&901&71&898&-7&8317&0.22$\pm$0.07&-0.09$\pm$0.01&6\\
    NGC 7142&326.29&65.782&3.1&2406&-628&2288&396&9255&0.00$\pm$0.04&-0.06$\pm$0.07&4\\
    NGC 7245&333.812&54.336&0.6&3210&-632&3145&-104&9507&-0.01$\pm$0.03&-0.09$\pm$0.03&5\\
    NGC 7762&357.472&68.035&2.0&897&-408&794&91&8784&0.05$\pm$0.07&-0.07$\pm$0.02&5\\
    NGC 7789&-0.666&56.726&1.5&2100&-901&1887&-196&9432&0.00$\pm$0.07&-0.07$\pm$0.02&4\\
    Ruprecht 171&278.012&-16.062&2.8&1522&1458&430&-82&6895&0.14$\pm$0.04&-0.05$\pm$0.03&6\\
    Skiff J1942+38.6&295.611&38.645&1.5&2378&700&2251&312&7964&0.10$\pm$0.06&-0.05$\pm$0.06&6\\
    UBC 3&283.799&12.326&0.1&1704&1214&1187&141&7223&-0.01$\pm$0.03&0.04$\pm$0.02&4\\
    UBC 6&344.01&51.187&0.7&1493&-387&1428&-199&8843&0.02$\pm$0.02&-0.08$\pm$0.03&6\\
    UBC 59&82.239&48.043&0.5&2585&-2439&789&334&10808&0.03$\pm$0.02&-0.02$\pm$0.06&3\\
    UBC 215&100.461&-5.243&0.4&1419&-1137&-842&-111&9514&0.08$\pm$0.05&-0.10$\pm$0.02&5\\
    \hline
    \end{tabular}
            \tablefoot{Positions and ages were computed by \citet{cantat-gaudin2020} from \textit{Gaia} DR2 \citep{Gaia2018}. Examples of abundances derived in the present work and the number of stars studied in each cluster are shown in the last columns of the printed table. The complete table with the LTE abundances of the 21 elements plus the NLTE of those available is published in the CDS version of the table.}
    \label{tab:ocs}
    \end{center}
    \end{table*}

    OCCASO project is dedicated to accurately measuring radial velocities and detailed elemental abundances in OCs, as tracers of the Milky Way disk (see Paper\,I for details). 
    To reach this goal, it is essential to obtain a sample as large and homogeneous as possible. To this end, OCCASO targets stars at the same evolutionary stage, the RC, to avoid star-to-star abundance variations caused by stellar evolution. RC stars can be easily identified even in sparsely populated color-magnitude diagrams. They are also brighter than main sequence (MS) stars, allowing us to observe them at further heliocentric distances. Since they are warmer than brighter giants, their spectra are less line crowded, which enables a more accurate determination of atmospheric parameters (effective temperature, surface gravity, and metallicity) and chemical abundances. This requirement constrains our sample to OCs older than 100\,Ma, since stars in younger systems have not had enough time to evolve to the RC stage. 

    In order to obtain representative values from the cluster's average abundances and radial velocities, we targeted at least four stars per cluster, although in some cases this number can reach up to 12 stars. We adopted this strategy because there is  a non-negligible probability of contamination from non-members, even when using membership determined from \textit{Gaia}'s proper motions and parallaxes. Moreover, some targets can be spectroscopic binaries, which complicates their analysis even being real cluster members.

        
        \begin{figure}
        \centering
        \includegraphics[width=\columnwidth]{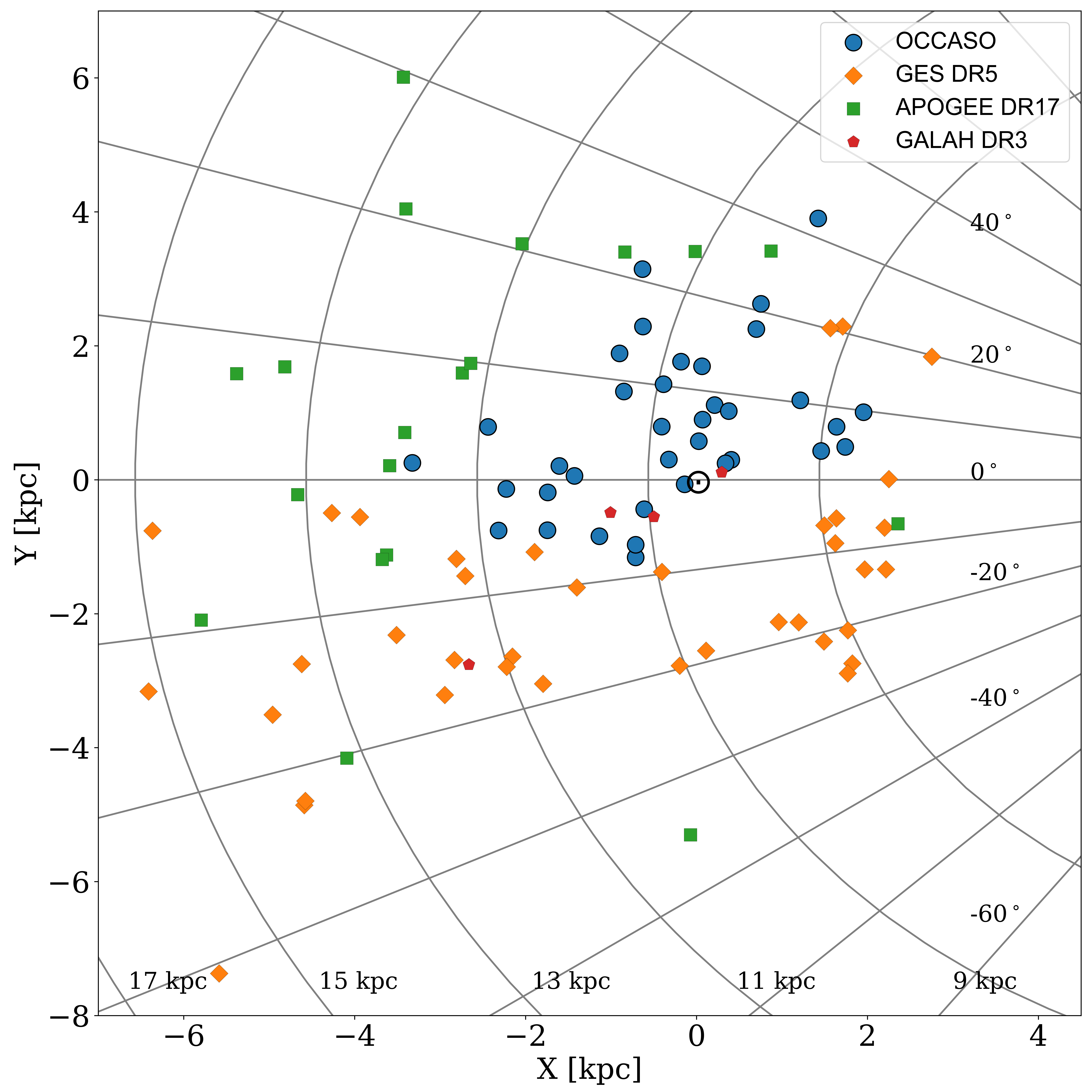}
        \caption{OCs observed in OCCASO (blue circles) together with OCs complying with similar requirements (at least four observed RC stars and spectral resolution above 20\,000) from GES DR5 (orange diamonds), APOGEE DR17 (green squares), and GALAH DR3 (red pentagons). All combined conform the OCCASO+ sample (see Sect. \ref{OPLUS}).}
        \label{fig:XY}
        \end{figure}

   OCCASO employs three high resolution spectroscopic facilities housed at Spanish observatories: the FIber-fed Echelle Spectrograph (FIES) at the Nordic Optical Telescope \citep[NOT, ][$R$$\sim$67\,000, 370<$\lambda$<900\,nm]{Telting2014}; the High Efficiency and Resolution Mercator Echelle Spectrograph (HERMES) at the Mercator telescope \citep[][$R$$\sim$82\,000, 377<$\lambda$<900\,nm]{Raskin2011}. Both FIES and HERMES are located at the Roque de los Muchachos Observatory in La Palma, Spain. The third instrument is the Calar Alto Fiber-fed Echelle spectrograph (CAFE) installed on the CAHA2.2m Telescope \citep[][$R$$\sim$62\,000, 400<$\lambda$<900\,nm]{caha}, situated at the Centro Astronómico Hispano-Alemán (CAHA) in Almería, Spain.

    The data reduction of the observed spectra is fully described in Paper\,IV. Briefly, the bias subtraction, flat-field correction, order tracing, extraction, and wavelength calibration were performed by the dedicated pipelines of each instrument: \texttt{HERMES DRS} for HERMES \citep{Raskin2011}; \texttt{FIEStool} for FIES \citep{Telting2014}; and \texttt{CAFExtractor} for CAFE \citep{Lill2020}. Our own OCCASO pipeline performs the sky and telluric line subtraction, combination of different exposures, normalization, and merging of the echelle orders. Additionally, we calculate the radial velocity of the stars by cross-correlation with a template, obtaining measurements with a precision of 10\,m\,s$^{-1}$. The improvements in the spectra combination procedure, described in Paper\,IV, have considerably reduced the noise of the final spectra, allowing for the detection of weaker lines than in our previous papers.

    We used the same sample described in Paper\,IV, where we discarded stars considered as non-members. Additionally, we rejected objects whose spectrum had a signal-to-noise (S/N) ratio lower than 50 pix$^{-1}$ to ensure a good quality abundance measurement. After this, our sample is formed by 194 stars belonging to 36 OCs, which main features are summarized in Table~\ref{tab:ocs}. Figure~\ref{fig:XY} shows the projection of the studied OCs onto the Galactic plane.

    \subsection{Methodology}\label{sec:methodology}    

    We followed a similar analysis strategy to that described in Papers\,II and\,III. Two different methods are used to compute the atmospheric parameters and chemical abundances: equivalent widths (EW) and spectral synthesis (SS). For the sake of homogeneity, in both cases, we use the \texttt{MARCS} atmosphere models \citep{Gustafsson2008} and the sixth version of the GES line list \citep{heiter2021}. We refer the reader to Paper\,II for details. 
   
    For the spectral synthesis, we used \ispec~\citep{bc2014,blancocuaresma2019}, and performed the analysis by comparison between the observed spectrum and synthetic one. The spectra were computed on the fly with the \texttt{SPECTRUM} radiative transfer code \citep{spectrum1994}. We varied the parameters to be measured until the chi-square value of the fit is minimized.   
    It provides effective temperatures, \teff, surface gravities, \logg, microturbulence velocities, \vmic, metallicities, \mh, and rotational velocities, \vsini.
    We consider that the line broadening is mainly due to rotation, by assuming a negligible macroturbulence velocity, as both broadening mechanisms are difficult to discern in red giants. \citep[e.g.,][]{thygesen2012}. The atmospheric parameters derived with SS are presented in Table \ref{tab:AP-ABU}.
        
    The equivalent widths were measured by \daospec, a code that detects and fits absorption lines \citep[][]{Stetson2008}. We took advantage of the \texttt{DOOp} wrapper \citep{cantat-gaudin2014}, which allowed us to batch-process the analysis and to determine the best \daospec~input parameters. The resulting EWs were fed to \texttt{GALA} \citep{GALA2013}, which derives the atmospheric parameters using the \texttt{WIDTH9} radiative transfer code \citep{Kurucz2005}, using the classical spectroscopic method based on Fe lines. We rejected the too weak and too strong lines outside the range -5.9 $\geq \log \frac{EW}{\lambda} \geq -4.8$. The atmospheric parameters derived with EW are \teff, \logg, \vmic, and $\mathrm{[M/H]}$ (Table~\ref{tab:AP-ABU}). Unlike the procedure carried out in Papers\,II and III, we did not perform a normalization with \daospec, since it was already done during the reduction process. Once the atmospheric parameters were determined by both methods, we compared them and determined the chemical abundances, as detailed in Sect. \ref{sec:spec_an}.

\subsubsection{Line list selection}
    \label{sec:line_list_selection}

    \begin{figure}
    \centering
    \includegraphics[width=\columnwidth]{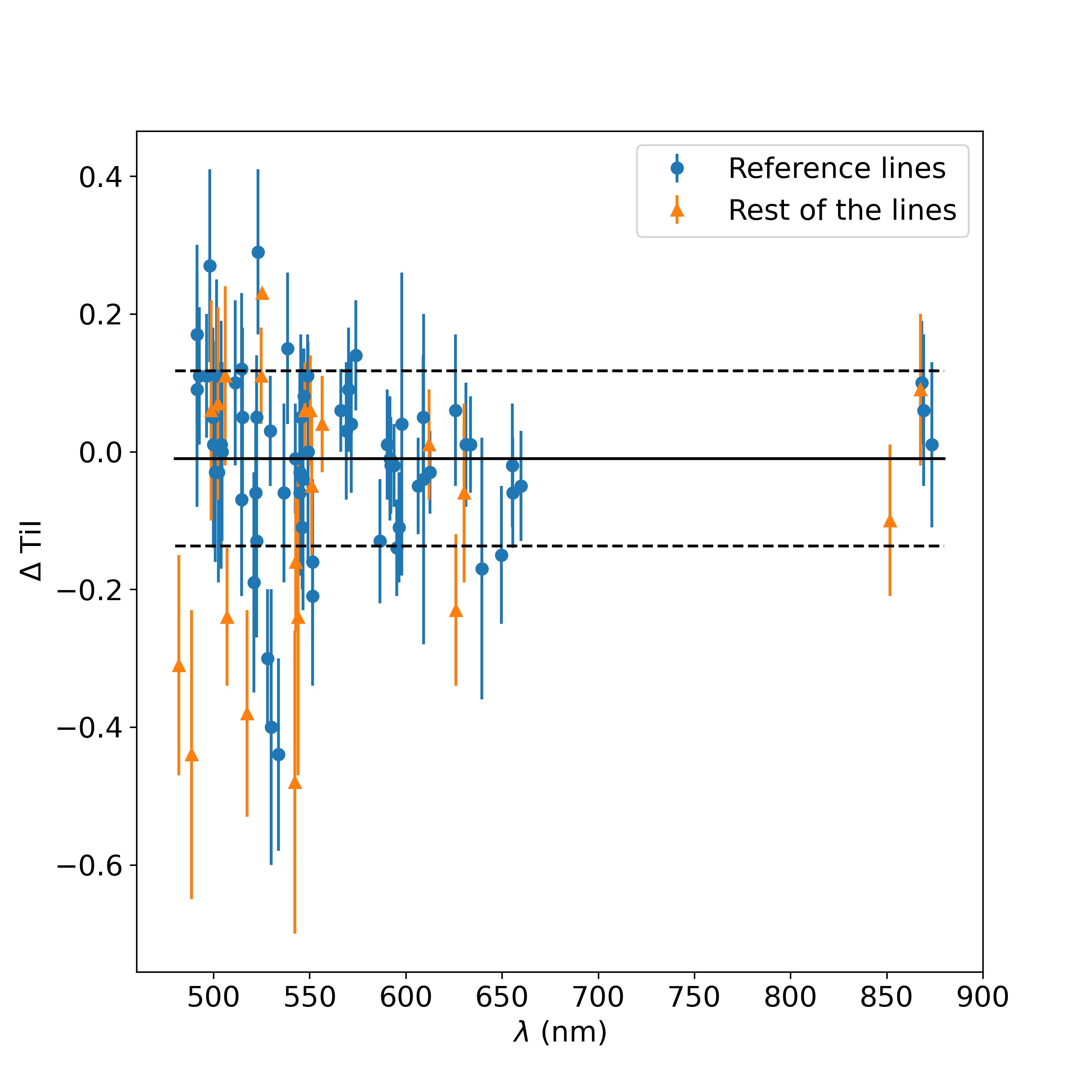}
    \caption{Difference between the abundance calculated using a given line and the \mbox{Ti\,{\sc i}} mean abundance represented versus the wavelength of the line. The blue circles are the reference lines, while orange triangles are the others. The distribution of the reference lines is used as selection criteria, keeping those lines that are in the region delimited by $\pm1\sigma$ (between dotted lines). Ti is used as an illustration; the procedure is the same for all chemical elements.}
    \label{fig_md}
    \end{figure}

    The sixth version of the GES line list is a compilation of atomic parameters of 80\,612 spectral lines \citep{heiter2021} covering the whole wavelength range of our spectra. The master line list provides two flags for each line: \emph{gf\_flag}, related to the reliability of the transition probabilities of the lines; and \emph{synflag} indicating whether the line is blended or not at the UVES resolution, $R$$\sim$47\,000. Both flags can take the values: "Y" for “yes, we recommend the use of this line;” "N" for “not recommended;” or "U" for “undecided” \citep[see][for details]{heiter2021}. However, we may be able to resolve some lines blended in UVES spectra as OCCASO uses a higher resolution ($R\geq$67\,000). Therefore, all the lines were potentially usable regardless of their \emph{synflag} values, as already suggested by \citet{heiter2021}.

    In the case of SS, which is less sensitive to line blends, we determined the atmospheric parameters using lines that lead to consistent abundances in a solar spectrum, as described in Paper\,II. 
    For the determination of atmospheric parameters with EW and abundances in both methods, we selected the most suitable lines with the following procedure: 
    First, we discarded all lines with \emph{gf\_flag}=N since they are considered of low accuracy. Second, we ran both methods on the whole sample, selecting lines which are measurable in at least 50 stars. 
    Third, we discarded the lines yielding different chemical abundances because of blends or inaccurate atomic parameters. 
    For each element, we calculated the difference between the abundance of each line and the mean abundance. By doing this for all stars, we obtained a distribution for each line, which we evaluated by calculating their median and standard deviation. The lines that have a dispersion higher than 0.25 dex are discarded. In Papers\,II and III, we employed a fixed value to discard lines with discrepant medians.
    On the contrary, in this work, we used reference lines in this process, considering as such those marked as \emph{gf\_flag}=Y and \emph{synflag}= Y or U. We evaluated the distribution of their medians and select all the lines in the $\pm1\sigma$ region (dashed lines in Fig. \ref{fig_md}). We repeated the process twice to improve the discarding of lines with discrepant values.    
    The EW and SS line lists are presented in Tables \ref{Tab:linelistEW} and \ref{Tab:linelistSS}, respectively.

    For the elements with less than ten detected lines (Na, Mg, Al, Zn, Sr, Y, Ba, Ce, Zr, and Nd), statistical criteria were not enough to assess the goodness of a line. In this case, we selected the best lines by visual inspection of our highest S/N spectra. Additionally, the Na lines at 589.00 and 589.59\,nm, labelled "YY," were discarded because they are contaminated by interstellar medium absorption lines.

\subsubsection{Impact of the initial guess values in \gala}
    \label{sec:GalaInit}
        
    {\gala} requires an initial guess of the atmospheric parameters to start the analysis. In Papers\,II and III, the same initial guess values were used to analyze the entire sample, \teff=4700\,K and \logg=2.50\,dex. In the present work, we investigate the impact of the initial guess values on the final result. To do so, we initialized the analysis with 21 different combinations of {\teff} and {\logg} in a grid covering 4100$\leq$\teff$\leq$5300\,K with a step of 200\,K, and 1.4$\leq$\logg$\leq$3.5\,dex with a step of 0.3\,dex. For each combination of initial guess values, we compute the atmospheric parameters of all stars.
         
    We used three independent criteria to evaluate the goodness of the results. First, the merit function provided by \gala, which estimates the quality of the global solution. It is calculated considering the values of the optimization parameters and the corresponding uncertainties \citep[see][for details]{GALA2013}. Second, we evaluated the difference between the derived atmospheric parameters and the initial guess values, selecting the results that satisfy $|T_{\rm eff,guess}-T_{\rm eff}|<$100\,K and $|\log g_{\rm guess}-\log g|<$0.5\,dex. This ruled out the solutions that converged to values very different from the initial guesses. Finally, we took into account the number of times that a result is obtained for the same star starting from the different guess values. The most repeated result may be the most reliable.

    We used the combination of the three criteria to choose the atmospheric parameters for each star. For 151 stars, 78\% of cases, the three criteria yield the same atmospheric parameters. These are the most robust EW results, flagged as 1 in column "GALAF" of Table \ref{tab:AP-ABU}. For 15\% of the cases, two of the criteria pointed to the same atmospheric parameters, while the other provides a different solution. In these cases, we selected the solution obtained from two of the criteria. They are flagged as 2 if the results were derived from the first and second criteria and flagged as 3, if it was obtained from the second and third criteria. There are no cases with agreement from the first and third criteria. Finally, in the remaining 13 stars, 7\% of the sample, every criterion provides a different solution. These cases are flagged as 4 and the adopted result is  the one with the best merit function.   
        
 \subsubsection{Solar abundance scale}
    
    To determine the solar abundances, we proceeded as in Papers\,II and III, retrieving the HARPS, UVES, and NARVAL solar spectra from the \gaia benchmark stars spectral library \citep{blancocuaresma2014}. We analyzed these spectra with the same methodology as the OCCASO ones and with the two methods, SS and EW. The solar abundances were calculated using the weighted mean of the values derived for each spectrum, with the standard deviation as uncertainty. The abundance values were calculated with the method chosen in Sect.~\ref{ChemAbu}. The derived values are listed in Table~\ref{tab:solar}. They are consistent, within twice the standard deviation, with the photospheric values obtained by \citet{Grevesse2007} and \citet{Asplund2009}. The only exceptions are Na, and Ba with differences of 0.2 and 0.4 dex respectively, being both below 3$\sigma$.

               \begin{table}[h!]
            \setlength{\tabcolsep}{1mm}
            \begin{center}
            \caption{Solar abundances calculated in this work, compared with \citet[][GAS07]{Grevesse2007} and \citet[][AGS09]{Asplund2009}.}
            \begin{tabular}{lcccc}
            \hline
            Element & This work & Method & GAS07 & AGS09 \\
             & [dex] & & [dex] & [dex] \\
            \hline
            \mbox{Fe\,{\sc i}}& 7.48$\pm$0.01 &EW& 7.45 $\pm$0.05&  7.50 $\pm$0.04 \\
            \mbox{Mg\,{\sc i}}& 7.50$\pm$0.01 &SS& 7.53$\pm$0.09 &  7.60 $\pm$0.04 \\
            \mbox{Si\,{\sc i}}& 7.43$\pm$0.01 &EW& 7.51$\pm$0.04  &  7.51 $\pm$0.03 \\
            \mbox{Ca\,{\sc i}}& 6.36$\pm$0.01  &SS& 6.31$\pm$0.04 &  6.34 $\pm$0.04  \\
            \mbox{Ti\,{\sc i}}& 4.91$\pm$0.02 &EW& 4.90$\pm$0.06 &  4.95 $\pm$0.05 \\
            \mbox{Na\,{\sc i}}& 6.49$\pm$0.03 &SS& 6.17$\pm$0.04 &  6.24 $\pm$0.04 \\
            \mbox{Al\,{\sc i}}& 6.55$\pm$0.02 &SS& 6.37$\pm$0.06  &  6.45 $\pm$0.03 \\
            \mbox{Sc\,{\sc ii}}& 3.18$\pm$0.02 &SS& 3.17$\pm$0.10 &  3.15 $\pm$0.04 \\
            \mbox{V\,{\sc i}}& 3.90$\pm$0.01 &SS& 4.00$\pm$0.02 &  3.93 $\pm$0.08 \\
            \mbox{Cr\,{\sc i}}& 5.59$\pm$0.02 &EW& 5.64$\pm$0.10 &  5.64 $\pm$0.04 \\
            \mbox{Mn\,{\sc i}}& 5.48$\pm$0.02 &SS& 5.39$\pm$0.03 &  5.43 $\pm$0.05 \\
            \mbox{Co\,{\sc i}}& 4.90$\pm$0.02 &SS& 4.92$\pm$0.08 &  4.99 $\pm$0.07 \\
            \mbox{Ni\,{\sc i}}& 6.27$\pm$0.02 &EW& 6.23$\pm$0.04 &  6.22 $\pm$0.04 \\
            \mbox{Cu\,{\sc i}}& 4.03$\pm$0.03 &SS& 4.21$\pm$0.04 &  4.19 $\pm$0.04 \\
            \mbox{Zn\,{\sc i}}& 4.35$\pm$0.02 &EW& 4.60$\pm$0.03 &  4.56 $\pm$0.05 \\
            \mbox{Sr\,{\sc ii}}& 3.03$\pm$0.07 &SS& 2.92$\pm$0.05 &  2.87 $\pm$0.07 \\          
            \mbox{Y\,{\sc ii}}& 2.08$\pm$0.04 &SS& 2.21$\pm$0.02 &  2.21 $\pm$0.05 \\
            \mbox{Zr\,{\sc ii}}& 2.58$\pm$0.04 &SS& 2.58$\pm$0.02 &  2.58 $\pm$0.04 \\
            \mbox{Ba\,{\sc ii}}& 2.19$\pm$0.02 &SS& 2.17$\pm$0.07 &  2.18 $\pm$0.09 \\       
            \mbox{Ce\,{\sc ii}}& 1.45$\pm$0.05 &SS& 1.70$\pm$0.10 &  1.58 $\pm$0.04 \\
            \mbox{Nd\,{\sc ii}}& 1.45$\pm$0.02 &SS& 1.45$\pm$0.05 &  1.42 $\pm$0.04 \\
                    \hline
            \end{tabular}
            \label{tab:solar}
            \end{center}
            \end{table}

\begin{figure}[hbt!] 
\centering
\includegraphics[width=\columnwidth]{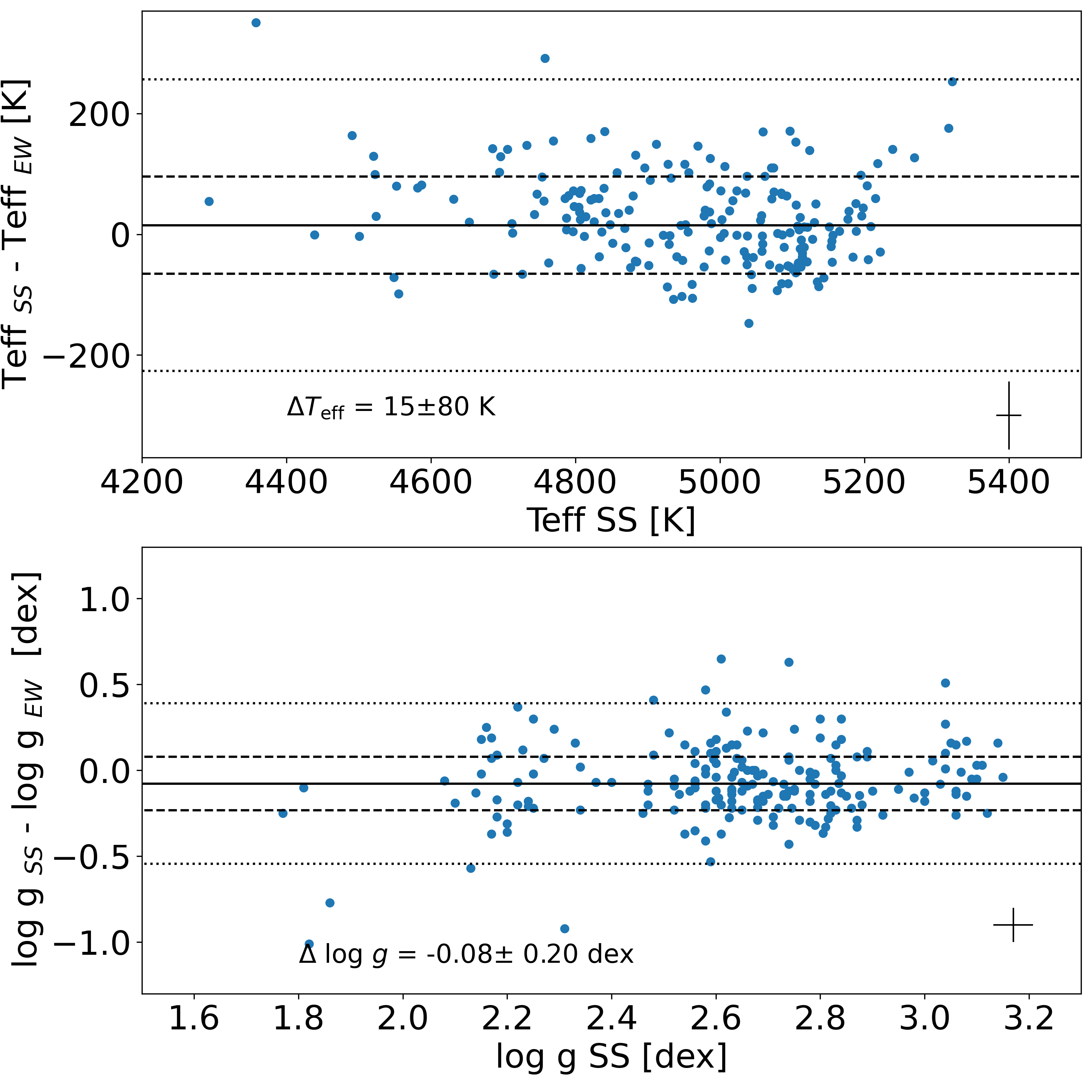}
\caption{Differences, in the sense SS-EW, of the {\teff} (top) and {\logg} (bottom) as a function of the SS values. Mean difference (solid line), standard deviation (dashed lines), and 3$\sigma$ levels (dotted lines), are shown in the plot. Typical error bars, calculated as the square of the quadratic sum of each method uncertainties, are shown in the bottom-right corner.} 
\label{fig:dif_var}
\end{figure}  

\begin{figure}[hbt!] 
          \includegraphics[clip,width=\columnwidth]{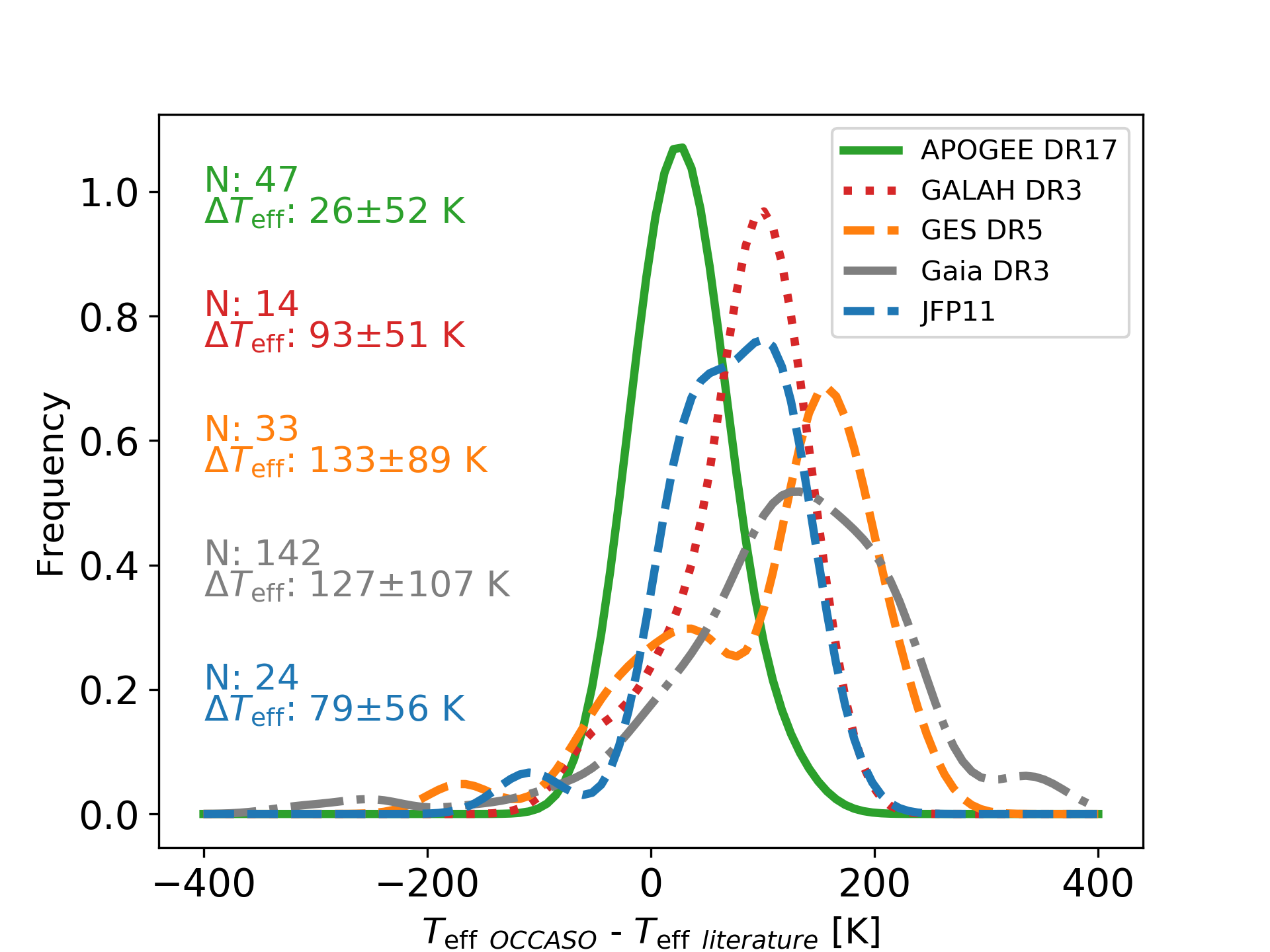}
          \includegraphics[clip,width=\columnwidth]{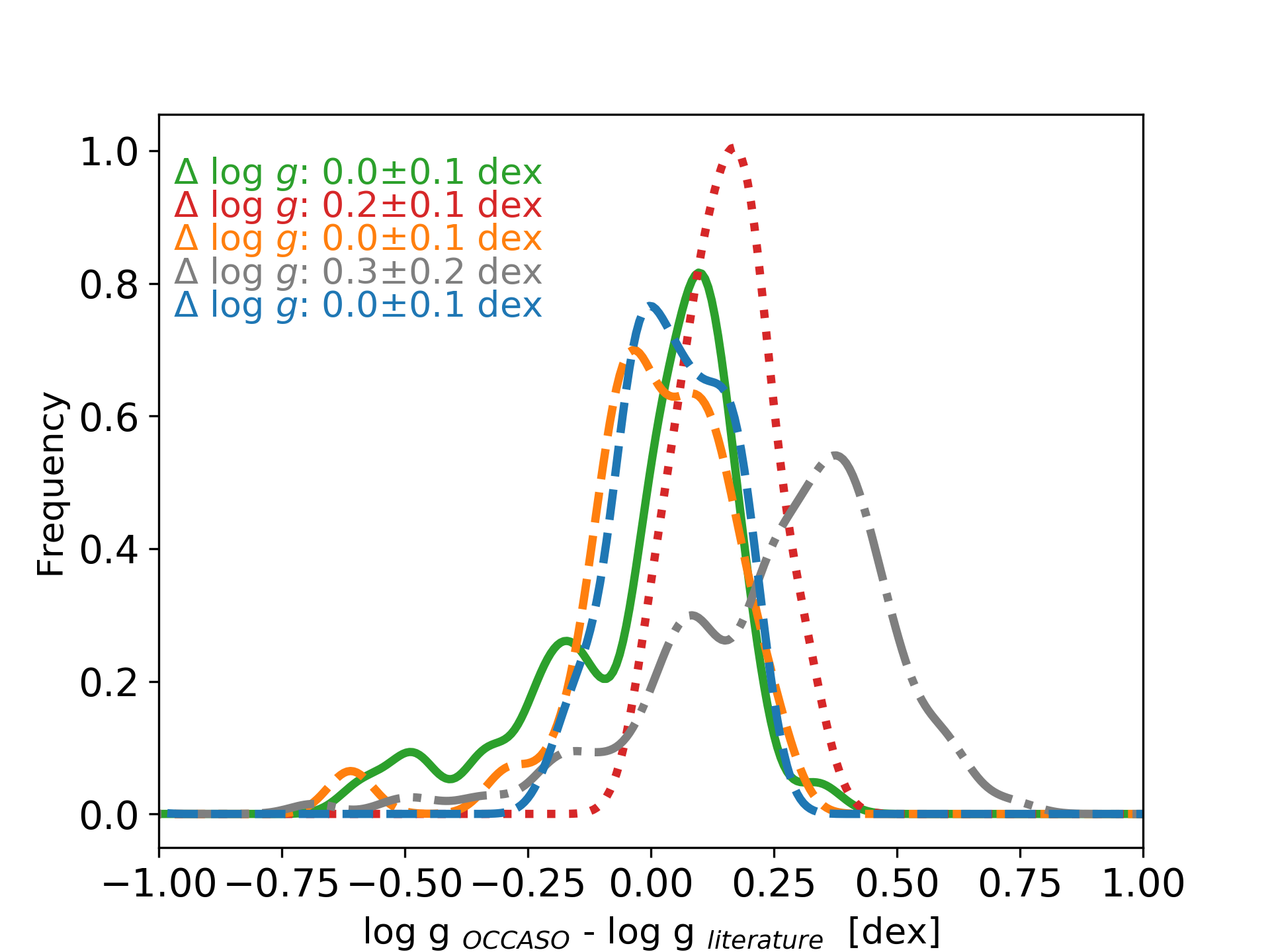}
        \caption{Distribution of the differences in {\teff} (top), {\logg} (bottom) between OCCASO and APOGEE\,DR17 (green), GALAH\,DR3 (purple), GES\,DR5 (orange), \gaia\,DR3 (grey) and \citet[][JFP, blue]{jacobson2011}. The histogram is smoothed. The mean of the differences and the standard deviation are shown in the panels.}
        \label{fig:biblio_AP}
        \end{figure}     

\section{Results of the spectroscopic analysis}\label{sec:spec_an}

\subsection{Atmospheric parameters}

              \begin{table*}[h!]
            \setlength{\tabcolsep}{1mm}
            \begin{center}
            \caption{Differences of the atmospheric parameters and abundances for the stars in common with the literature, in the sense ours-others.}
            \begin{tabular}{lcccccccccc}
            \hline
                Parameter & APOGEE DR17&$N$& GALAH DR3&$N$& GES DR5&$N$&{\gaia} DR3&$N$&\citet{jacobson2011} &$N$\\
            \hline

        $T$\textsubscript{eff} [K] &  26.0$\pm$52& 49& 93.0$\pm$51& 14& 133.0$\pm$89& 35 &127$\pm$107&142& 79$\pm$56 & 24\\   
        $\log g$[dex]& 0.0$\pm$0.1& 49& 0.2$\pm$0.1& 14& 0.0$\pm$0.1& 35 &0.3$\pm$0.2&142& 0.0$\pm$0.1 & 24 \\      
        $[$Fe/H$]$ [dex]& 0.0$\pm$0.1& 49& 0.0$\pm$0.1& 14& 0.0$\pm$0.1& 35&-0.1$\pm$0.1&142&0.0$\pm$0.1& 24   \\ 
        $[$Mg/Fe$]$ [dex]& 0.0$\pm$0.1& 47& 0.0$\pm$0.1& 14& -0.1$\pm$0.1& 33& & &-0.2$\pm$0.1& 24  \\ 
        $[$Si/Fe$]$ [dex]& 0.1$\pm$0.1& 49& -0.1$\pm$0.1& 14& 0.0$\pm$0.1& 35&  & &-0.1$\pm$0.1& 24  \\ 
        $[$Ca/Fe$]$ [dex]& 0.0$\pm$0.1& 49& -0.1$\pm$0.1& 14& 0.0$\pm$0.1& 35&  & &0.1$\pm$0.1& 24  \\ 
        $[$Ti/Fe$]$ [dex]& 0.1$\pm$0.1& 49& 0.0$\pm$0.1& 14& 0.1$\pm$0.2& 35&  & &0.2$\pm$0.1& 24  \\ 
        $[$Na/Fe$]$ [dex]& 0.1$\pm$0.1& 47& 0.0$\pm$0.1& 14& 0.1$\pm$0.1& 33&  & &0.0$\pm$0.1& 24  \\  
        $[$Al/Fe$]$ [dex]& 0.2$\pm$0.1& 47& 0.0$\pm$0.1& 14& 0.0$\pm$0.1& 33&  & &-0.2$\pm$0.1& 7  \\
        $[$V/Fe$]$ [dex]& 0.1$\pm$0.2& 47& -0.1$\pm$0.1& 12& 0.0$\pm$0.1& 33&&& -- & 0  \\ 
        $[$Cr/Fe$]$ [dex]& 0.0$\pm$0.1& 49& 0.0$\pm$0.1& 14& 0.0$\pm$0.1& 35&  & &0.0$\pm$0.1& 7  \\         
        $[$Mn/Fe$]$ [dex]& -0.1$\pm$0.1& 47& -0.2$\pm$0.1& 14& 0.0$\pm$0.1& 33&&& -- & 0  \\ 
        $[$Co/Fe$]$ [dex]& 0.0$\pm$0.1& 47& 0.1$\pm$0.1& 14& 0.1$\pm$0.2& 33&&& -- & 0  \\ 
        $[$Ni/Fe$]$ [dex]& 0.0$\pm$0.1& 49& 0.0$\pm$0.2& 13& 0.0$\pm$0.1& 35&&&  0.1$\pm$0.1& 24  \\
        $[$Cu/Fe$]$ [dex]& -- & 0& 0.0$\pm$0.1&  14& 0.0$\pm$0.2& 32&&& -- & 0  \\ 
        $[$Zn/Fe$]$ [dex]& -- & 0& -0.1$\pm$0.2&  13& 0.0$\pm$0.1& 33&&& -- & 0  \\
        $[$Y/Fe$]$ [dex]& -- & 0& -0.2$\pm$0.2&  14& 0.1$\pm$0.1& 33&&& -- & 0 \\
        $[$Zr/Fe$]$ [dex]& -- & 0& -0.1$\pm$0.1&  11& -0.1$\pm$0.3& 21&&& -0.1$\pm$0.1 & 7 \\        
        $[$Ba/Fe$]$ [dex]& -- & 0& -0.3$\pm$0.2&  14& 0.1$\pm$0.1& 31&&& -- & 0 \\ 
        $[$Ce/Fe$]$ [dex]& 0.1$\pm$0.2& 47& -- &  0& 0.1$\pm$0.2& 32&&& -- & 0 \\ 
        $[$Nd/Fe$]$ [dex]& -- & 0& -0.2$\pm$0.0&12& -0.1$\pm$0.1& 33&&& -- & 0 \\

                    \hline
            \end{tabular}
                \tablefoot{We only consider those works with more than ten stars in common. $N$ is the number of stars in common with OCCASO per study and parameter. Flags of surveys used to select stars for comparison: APOGEE DR17: ASPCAPFLAG = 0 or 4; GES DR5: SFLAGS $\neq$ SNR,SRP,NIA; GALAH DR3: flag\_sp = 0, flag\_fe\_h=0; Gaia DR3 \& flags 1-7, 10 and 12 = 0, flag 8 and 9 $\leq$ 1.}
            \label{tab:lit_dif}
            \end{center}
            \end{table*}


The effective temperatures and surface gravities, derived by each method with their uncertainties, are listed in Table \ref{tab:AP-ABU} and compared in Fig.~\ref{fig:dif_var}. The {\teff} obtained with SS are slightly higher by 15$\pm80$\,K than those derived with EW. The typical uncertainties for EW and SS are 55\,K and 17\,K, respectively. For surface gravity, we find that EW yields values 0.08\,dex higher than SS. The standard deviation of the differences is 0.2\,dex, which is larger than the average uncertainties for EW, 0.09\,dex, and SS, 0.04\,dex, respectively. This difference and the dispersion may be explained by the well known difficulties in deriving {\logg} from spectroscopy, even from high quality and large wavelength coverage spectra as in our case. These results are compatible with the values obtained in Paper\,II for a smaller sample. Other studies using different methods find similar differences among methods \citep[e.g.,][]{Smiljanic2014}.

    
Owing to the fact that {\teff} and {\logg} values derived from both methods are compatible within the uncertainties, we obtained the weighted average of them, also listed in Table~\ref{tab:AP-ABU}, following the same strategy as in Paper\,II. Because general metallicity and microturbulence velocity are modelled differently in both methods, we do not attempt to combine them. The average values of {\teff} and {\logg} are used to calculate again the microturbulence and overall metallicity in both methods independently, and additionally, the rotational velocity in SS. This procedure decreases the differences between the two methods in the chemical abundance determination (see Paper\,II). The {\teff} and {\logg} uncertainties of each method were added in quadrature to calculate the uncertainties of the average values. 

Our sample includes several stars that have also been observed by other spectroscopic surveys. We compared our sample with studies that share a minimum of 10 stars, as shown in Fig.~\ref{fig:biblio_AP} and listed in Table~\ref{tab:lit_dif}: APOGEE DR17 \citep{APOGEEDR17}, GALAH DR3 \citep{GALAHDR3}, GES DR5 \citep{Randich2022}, \textit{Gaia} DR3 GSP-Spec \citep{GC_recioblanco2022} and \citet{jacobson2011}. In all cases, star selection was based on different flags provided by each survey, which are described in Table~\ref{tab:lit_dif}.

There is an excellent agreement with APOGEE DR17 within the uncertainties. For GALAH DR3, we obtained slightly larger values, although this comparison is based on only 14 stars. Moreover, the distribution in {\teff} shows a tail towards negative differences. The difference with {\teff} from GES DR5 shows a double peaked distribution, which yields a large standard deviation of 89\,K. The highest peak shows that our {\teff} are higher than those provided by GES DR5. In contrast, their {\logg} values show a good agreement with ours. The comparison with {\gaia} DR3 shows wider distributions, as expected from the larger uncertainties of the \gaia measurements, and our values of {\teff} and {\logg} are larger. The irregular distribution obtained for the differences in {\logg} reflects the greater difficulty in deriving gravities from spectroscopy in the small wavelength range covered by \textit{Gaia}. We performed the comparison after applying the {\logg} correction proposed by \citet{recioblanco2022}. In the case of \citet{jacobson2011}, our {\teff} are slightly higher than theirs, but {\logg} values show a good agreement. We have also compared our results with other high-resolution studies ($R\sim$ 20\,000) with less than ten stars in common. There are no clear systematics with any of those studies.

\subsection{Stellar chemical abundances}\label{ChemAbu}  

    From SS, we determined abundances for 21 chemical elements: Fe, Mg, Si, Ca, Ti, Na, Al, Sc, V, Cr, Mn, Co, Ni, Cu, Zn, Sr, Y, Zr, Ba, Ce, and Nd. Additionally, for those elements with a significant number of unblended lines and not significantly affected by hyperfine or isotopic structure, we also derived abundances from EW for Fe, Cr, Ni, Ti, Si, Ca, and Zn.
    
    The abundances were calculated as the weighted average of the values derived for each individual line. The abundance uncertainties were computed as $\frac{\sigma}{\sqrt{n}}$, being $\sigma$ the standard deviation and $n$ the number of lines. We took into account the low statistic correction factor in $\sigma$ by applying the equation~5 of \citet{Roesslein2007}. When only one line was measured (Zn, Sr, Zr, and Ce), we adopted the line abundance error as the elemental uncertainty. The abundances computed by both methods are listed in Table \ref{tab:AP-ABU}.
    
    Several stars in our sample were observed with more than one instrumental configuration. This is used to check the consistency of our results. 
    For Fe abundances, we find mean differences of 0.03$\pm$0.04\,dex in EW, and 0.02$\pm$0.03\,dex in SS, in good agreement with the quoted uncertainties. The situation is similar for other elements. Therefore, in the case of stars observed with more than one instrumental configuration, the final abundances were calculated as the weighted mean of the several values obtained. We found correlations with temperature for the elements Si, Na, Al, V, Ni, Sr, Y, Zr, Ba, Ce and Nd having slopes, in absolute values, greater than 1.5x10$^{-4}$ dex/K. This is because in our stars temperature and age are correlated; therefore, as the abundances of these elements depend on age (see Sect. \ref{sec:age_gradient}), they will also exhibit a temperature dependence. 
      
\begin{figure}[hbt!] 
    \centering
    \includegraphics[width=\columnwidth]{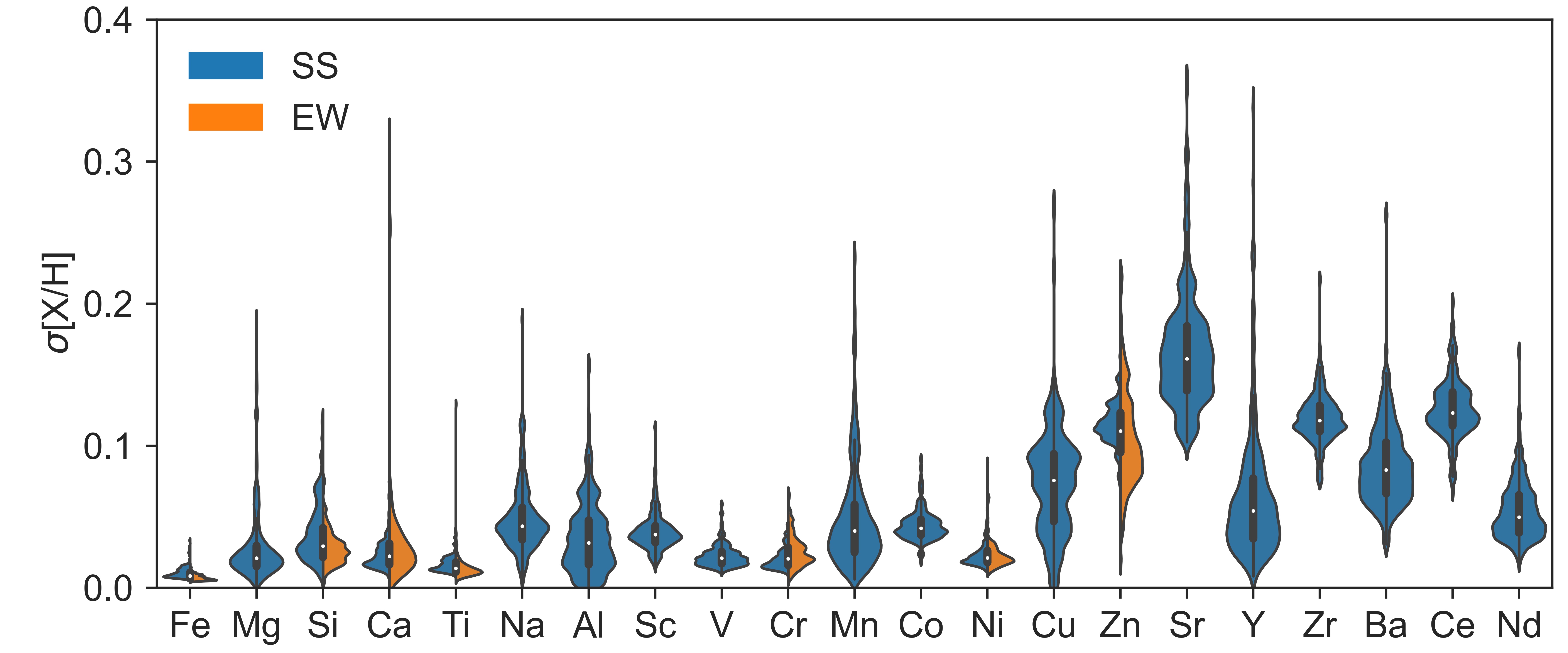}
    \caption{Distribution of the stellar [X/H] uncertainties of the 21 chemical elements derived with SS or EW methods.}
    \label{fig:SS_abu_uncer}
    \end{figure}

    Figure~\ref{fig:SS_abu_uncer} shows the distributions of the [X/H] uncertainties of the 21 elements measured in the stars of the sample.
    The elements from Fe to Ni have mean uncertainties of about 0.05\,dex, while from Cu on they have slightly higher uncertainties $\sim$0.1\,dex. This is explained by the small number of lines measured for these elements. Sr has the most imprecise measurements, with mean uncertainties of 0.15\,dex.
    
    \begin{figure}[hbt!] 
    \centering
    \includegraphics[width=\columnwidth]{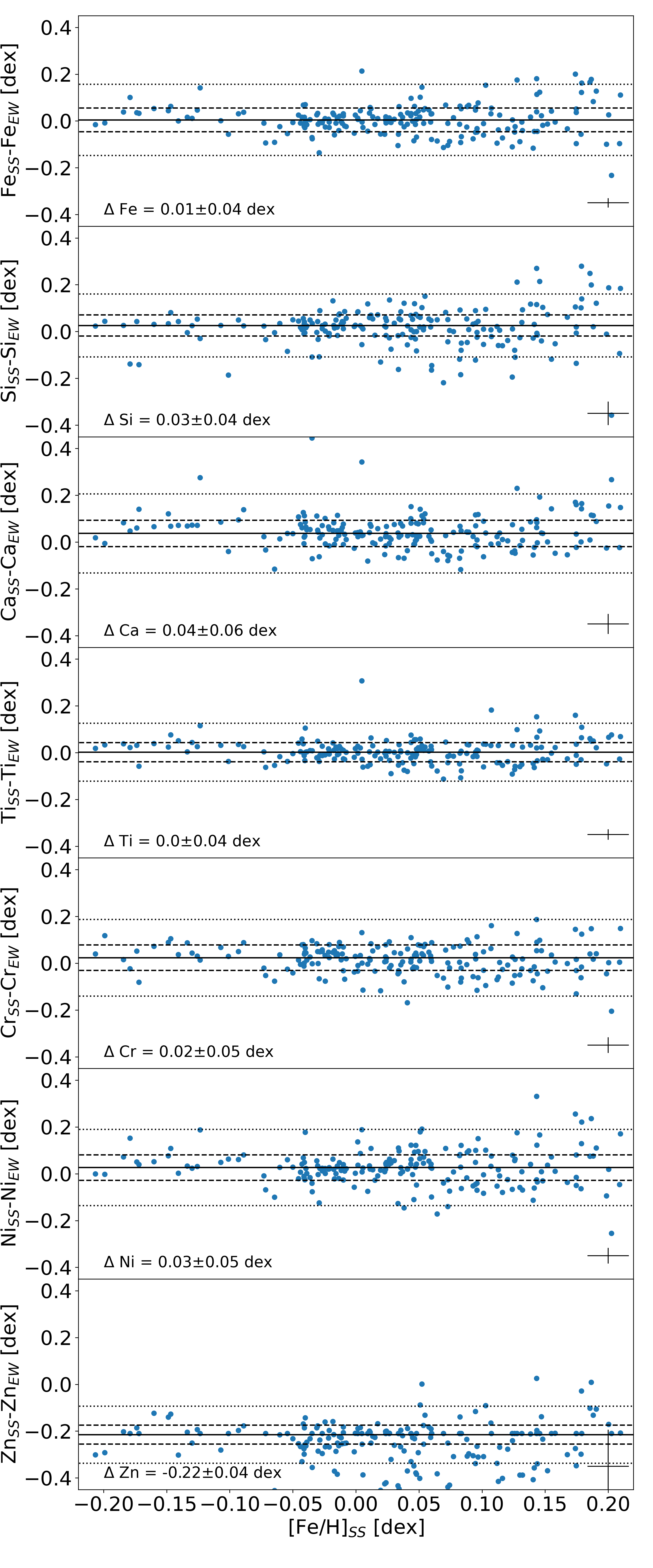}
    \caption{Differences of the abundances computed by both EW and SS methods versus SS [Fe/H] values. The mean differences and standard deviations are shown in each panel. In the bottom-right corner of each panel, we plot the mean uncertainty.}
    \label{fig:dif_var_Fe}
    \end{figure}    
    
         In Fig.~\ref{fig:dif_var_Fe}, we compare the derived abundances of the seven elements studied with the two methods. We find good agreement between methods except in the case of Zn for which, there is a systematic difference of $0.22\pm0.04$\,dex. The origin of the difference is unknown. It is not due to the radiative transfer code used in SS since we obtain similar results when we repeat the analysis using Turbospectrum \citep{Plez2012}. The mean difference for Fe abundance is 0.01\,dex, with a standard deviation of 0.04\,dex. So, we no longer find any difference in Fe abundance found in Paper\,II of 0.07$\pm$0.05\,dex. This may be explained by the improvements in the continuum placement implemented in Paper\,IV, and the improved EW analysis discussed above.
         The stars that have the largest differences between their abundances measured by the two methods are those with the lowest S/N. 
         When the abundance of an element is measured by both methods, we evaluate the number of used lines, the abundance internal error, and systematic differences with the literature to select which method we rely on the most. We adopt the SS abundances except for Fe, Si, Ti, Cr, Ni, and Zn, for which we use the EW determinations.

         Table~\ref{tab:lit_dif} shows the differences with the literature for the studied elements. Only the works with at least ten stars in common with our sample have been included. For [Fe/H], there is an excellent agreement with APOGEE DR17, GES DR5, and GALAH DR3. \gaia DR3 shows a systematic difference of $-0.1\pm0.1$\,dex. We perform this comparison after applying the correction in [Fe/H] proposed by \citet{recioblanco2022}. 
        We do not find general systematic differences with the literature for the rest of the elements. Furthermore, we have compared our results and other high-resolution studies ($R\gtrsim$20\,000) that have ten or more stars in common with our sample. Our analysis reveals no evident patterns or consistent discrepancies between these studies and ours.

    \subsection{Cluster chemical abundances}
    \label{Cluster chemical abundances}

        We adopt the membership analysis performed in Paper\,IV based on the proper motions and radial velocities. There are 32 stars reported in the literature as spectroscopic binary members (see Paper\,IV for details). We checked for the presence of double lines in their spectra using \ispec, but we do not find signs of secondary stars in any of them. We consider their spectra valid for abundance analysis.
        
\begin{figure}[hbt!] 
        \centering
        \includegraphics[width=\columnwidth]{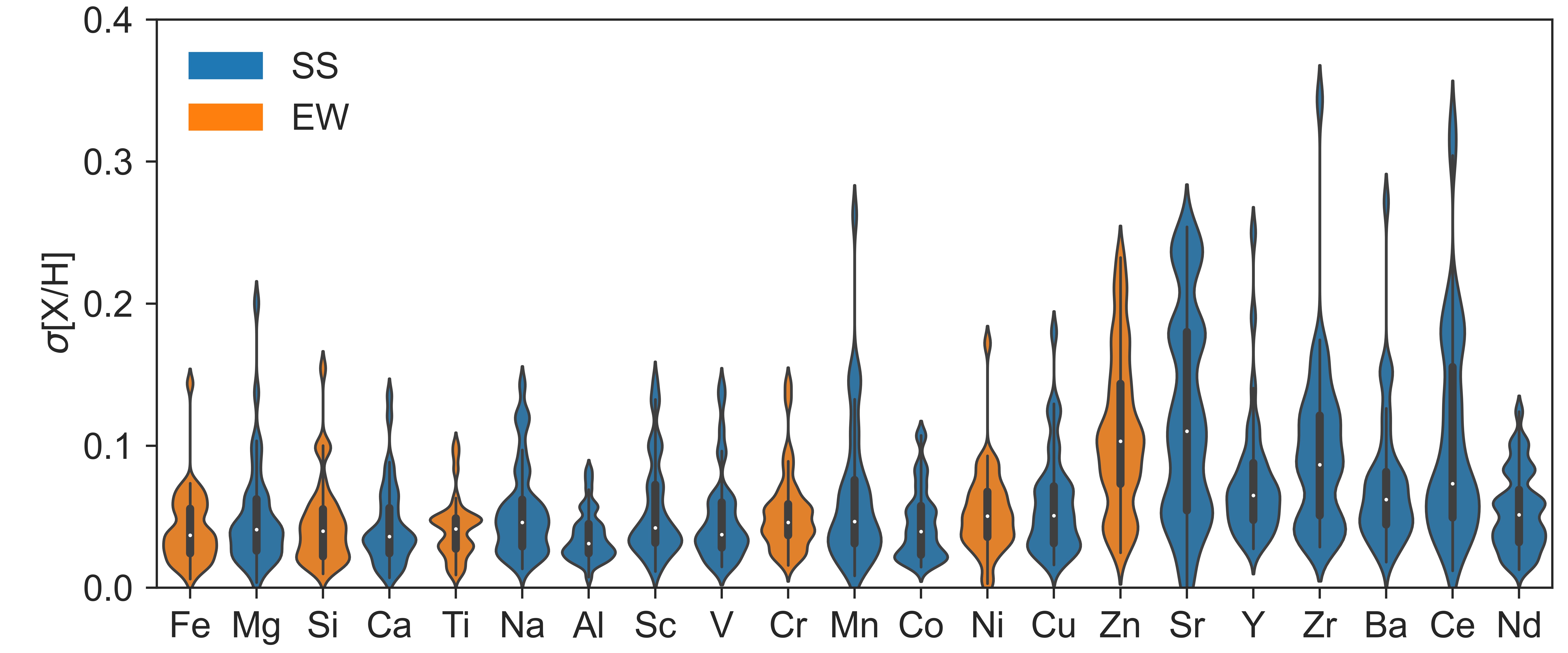}
        \caption{Distribution of the [X/H] standard deviations derived with the SS (blue) and EW (orange) methods for the 36 OCs in our sample.}
        \label{fig:SSOC_abu_uncer}
        \end{figure}

        The abundance of each cluster is the weighted average of those of its member stars. The associated error is then the standard deviation of the abundances multiplied by the low statistics correction factor (see Sect.~\ref{ChemAbu}). The full cluster abundances are listed in the full version of Table \ref{tab:ocs}, available at the CDS. In Fig.~\ref{fig:SSOC_abu_uncer}, we show the distribution of standard deviations for each element.
        For the elements from Fe to Cu, the mean standard deviation is around 0.05\,dex, and it is higher for the elements from Zn to Nd. This is a direct consequence of the higher uncertainties measuring those elements. NGC~6791 shows the largest standard deviations because it is the most distant cluster in the sample and has been measured with the lowest S/N.   
        
 \begin{figure}
        \centering
        \includegraphics[width=\columnwidth]{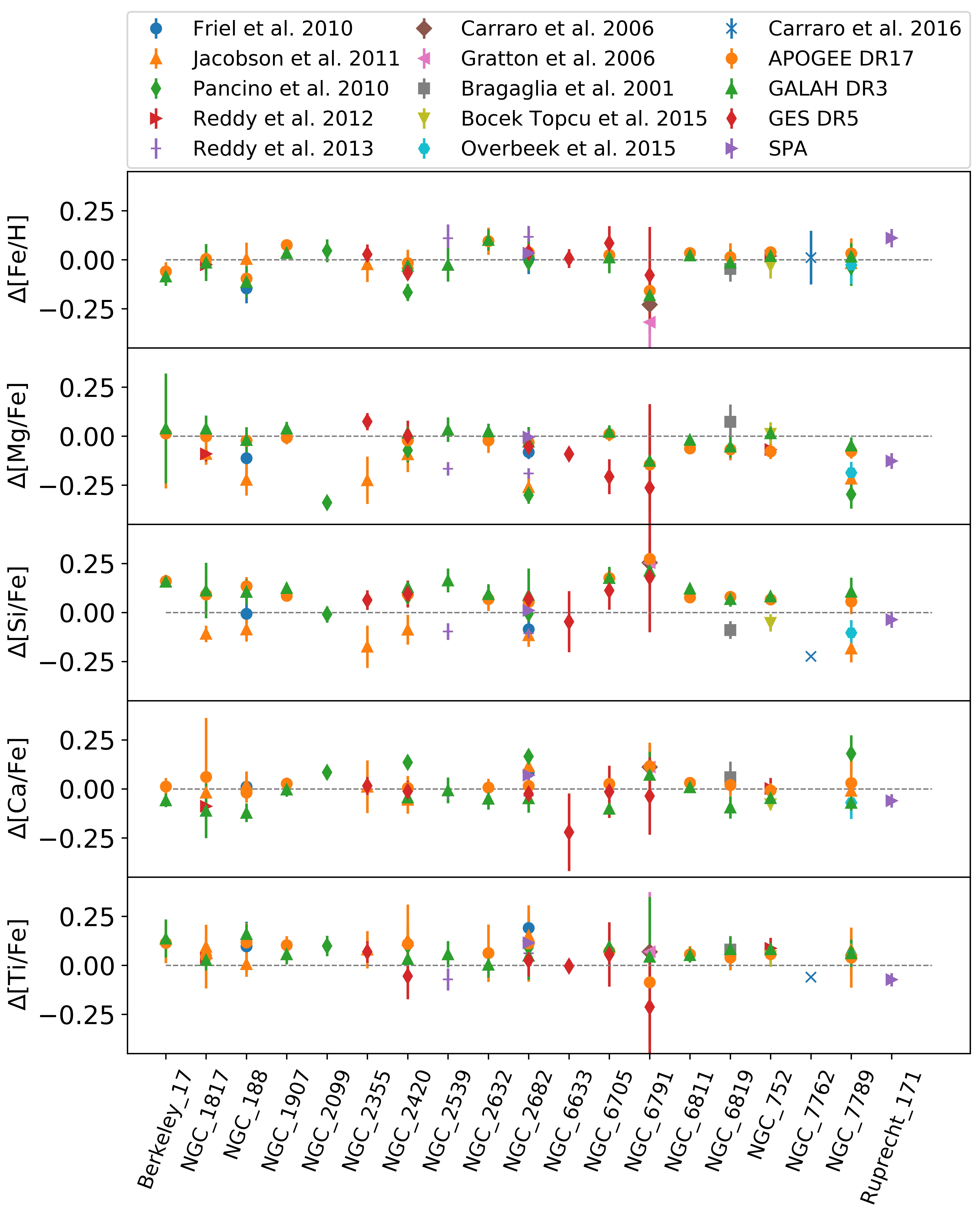}
        \caption{Comparison of the abundances of Fe and $\alpha$ elements of the OCs in our sample with the literature, in the sense this work minus literature.}
        \label{fig:litOC}
        \end{figure}

    In Fig. \ref{fig:litOC} we compare our abundances of Fe and $\alpha$ elements with other high-resolution studies ($R > $20\,000). In Fig.~\ref{fig:litOC_annex} we show the comparisons for the rest of the elements. There is a general good agreement with the literature. The differences are at the same level than the ones found in the star-by-star comparison in Table \ref{tab:lit_dif}. The OC that presents the most considerable difference with the literature is NGC~6791, specially in [Fe/H], although the values of some studies are compatible with ours considering the errors (e.g., GES DR5). This was already noticed and reported in Paper\,III.

    \subsection{Nonlocal thermodynamic equilibrium corrections}
    \label{NLTE}
    
    Some of the elements could be affected by nonlocal thermodynamic equilibrium (NLTE), so we have reviewed the effect of considering it in our sample. We calculated the corrections for Fe, Mg, Si, Ca, Ti, Cr, Mn, and Co using Spectrum Tools\footnote{https://nlte.mpia.de/index.php} that make use of the works of \citet{Bergemann2012}, \citet{Bergemann2015}, \citet{Bergemann2013}, \citet{Mashonkina2007}, \citet{Bergemann2011}, \citet{Bergemann2010b}, \citet{Bergemann2008} and \citet{Bergemann2010}, respectively. We corrected Ba from the results of \citep{Korotin2015} and Na based on \citep{Lind2011}. 
    
\begin{figure}[hbt!] 
        \centering
        \includegraphics[width=0.9\columnwidth]{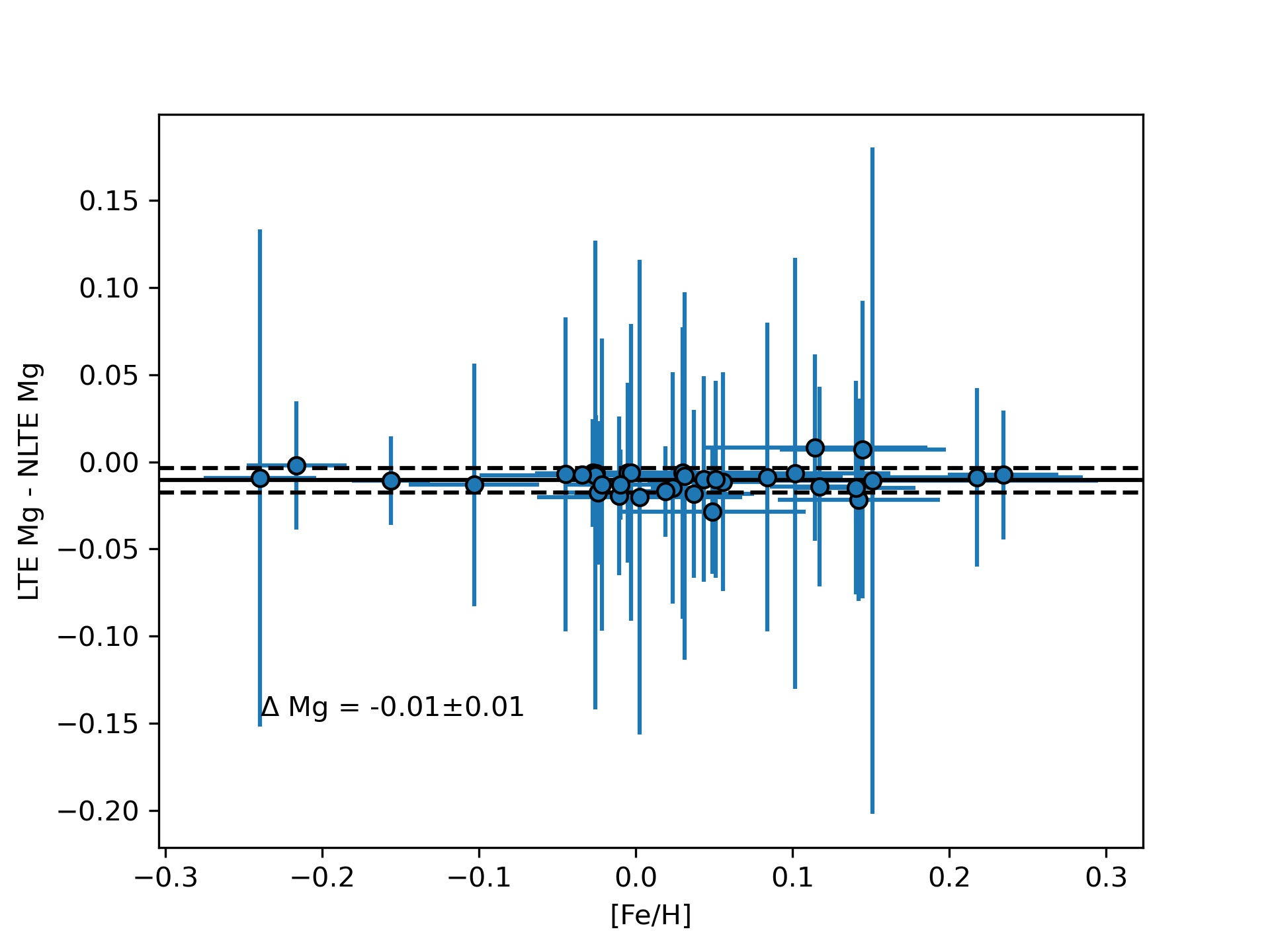}
        \caption{LTE-NLTE comparison per OC for Mg.}
        \label{fig:Mg}
        \end{figure}

\begin{figure}[hbt!] 
        \centering
        \includegraphics[width=0.9\columnwidth]{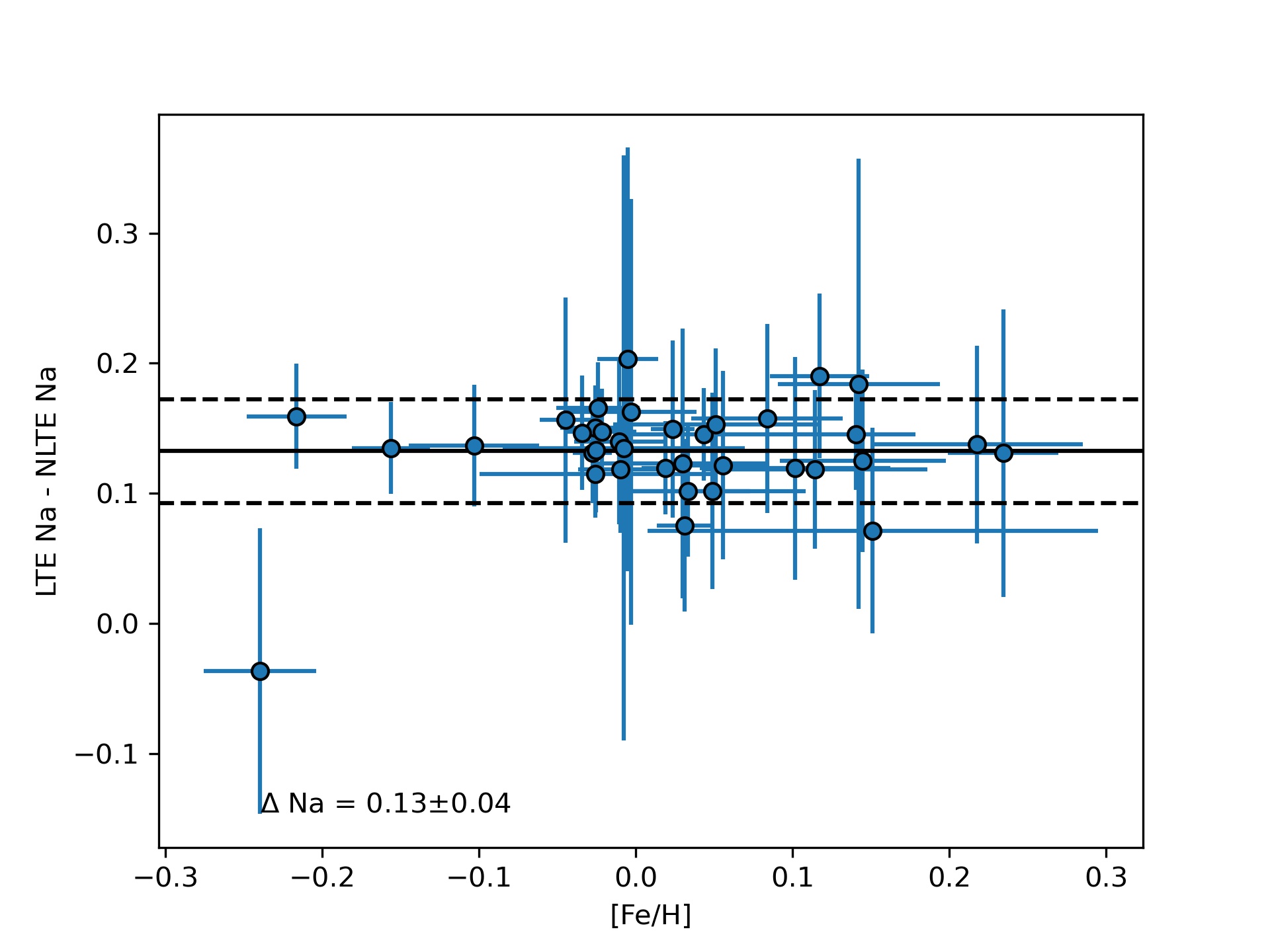}
        \caption{LTE-NLTE comparison per OC for Na.}
        \label{fig:Na}
        \end{figure}

    When comparing the results considering local thermodynamic equilibrium (LTE) with NLTE, we find that Fe, Mg (Fig. \ref{fig:Mg}), Si, Ca, Ti, Mn, and Ba do not show significant differences when applying NLTE corrections. Cr and Co have slightly higher values when applying NLTE, finding LTE-NLTE differences of -0.06$\pm$0.02 dex and -0.03$\pm$0.02 dex, respectively. While Na has systematically lower values when correcting for NLTE (Fig. \ref{fig:Na}) with differences LTE-NLTE of 0.13$\pm$0.04 dex. 
    The abundance values per OC applying NLTE are published in the full version of Table~1, available at the CDS.
        

\section{Galactic trends} 
\label{OPLUS}
    We use the mean abundances of our OCs sample in combination with the positions and ages from \citet{cantat-gaudin2020}, and the orbital parameters derived in Paper\,IV to investigate the abundance trends in the Galactic disk.

    The OCCASO sample provides high-precision abundances derived homogeneously, but it is limited in Galactocentric radius to $R_{\rm GC} < 11.7$\,kpc and Galactic azimuth  $\phi<$ -10\degr~ (Fig.~\ref{fig:XY}). In order to enlarge the spatial coverage, we created OCCASO+ adding OCs  from high-resolution ($R$>20\,000) surveys: GES DR5 \citep{Magrini2022}, APOGEE DR17 \citep{myers2022}, and GALAH DR3 \citep{spina2021}. The OCCASO and GES samples complement each other due to their similar spectral resolutions and wavelength coverage. Additionally, each one observes a distinct hemisphere, resulting in different Galactic azimuths being studied. APOGEE DR17 covers 0\degr $\leq$ $\phi$ $\leq$ 30\degr and $R_{\rm GC} > 11$~kpc not covered by OCCASO due to its limiting magnitude. 
    
    In all cases, we selected OCs that have a minimum of four stars studied in the RC region. For those OCs observed  by more than one survey, we prioritized the results with the highest resolution in the following sequence: OCCASO, GES, and APOGEE. Owing to the small number of systems sampled by GALAH, we only selected them if they had not been observed by any of the others. As the other works we compare with mostly use LTE values, we use the OCCASO LTE values in this section.     
    We discarded the V and Co abundances from \citet{myers2022} due to their large uncertainties in comparison with OCCASO. In total, OCCASO+ contains 99 OCs: 36 from OCCASO, 40 from GES DR5, 19 from APOGEE DR17, and 4 from GALAH DR3 (Fig.~\ref{fig:XY}). This means that this sample matches 64\% with the GES23 sample, 20\% with the high-quality sample from \citet{myers2022}, and 17\% with the sample of \citep{spina2021}.  
    We  analyzed the stars of OCs in common between the different surveys and OCCASO by calculating the differences of [X/Fe] values. We did not find any dependence of the abundance differences on atmospheric parameters or [Fe/H]. What we did find are slight [X/Fe] abundances zero points between studies (Fig. \ref{fig:litOC_annex}). We applied the abundance offsets to the literature samples so that all abundances are on the same scale.   
    Special attention should be paid to the GALAH abundances, as several of the elements are published with NLTE calculations \citep{GALAHDR3}. By checking the stars in common (Table \ref{tab:lit_dif}), we did not find dependencies with atmospheric parameters nor zero points higher than with other surveys. The only exception is [Ba/Fe], for which we find differences around 0.4 dex. NLTE cannot explain these differences (see Sect. \ref{NLTE}) so the difference between GALAH and OCCASO results must be due to other sources. We have, therefore, used the GALAH abundances in OCCASO+.
    The way in which OCCASO+ has been selected makes it the most complete sample of OCs with precise abundances.   
    In the following subsections, we analyze different Galactic disk chemical abundance trends with both OCCASO and OCCASO+ samples.

    \subsection{Abundance dependence on [Fe/H]}
        \label{sect:metallicity}

        \begin{figure*}
        \centering
        \includegraphics[width=18.5cm]{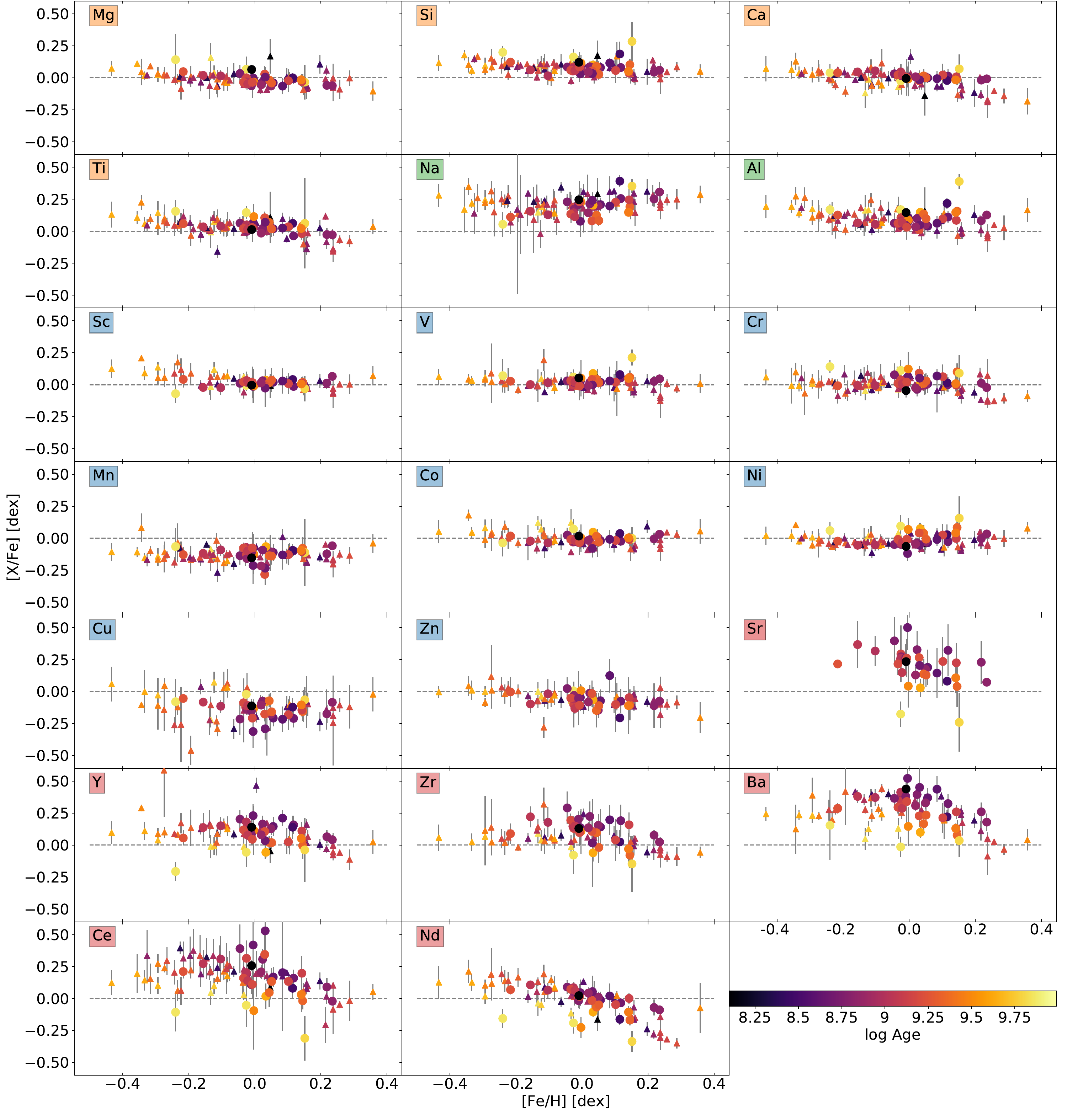}
        \caption{Abundance [X/Fe] ratios as function of [Fe/H] for OCCASO (circles) and the OCCASO+ (triangles) samples, respectively, color-coded with the age of the OCs. The color in the name of the element indicates the nucleosynthetic group: $\alpha$ (orange), odd-Z (green), Fe-peak (blue), and neutron-capture (pink).}
        \label{fig:Abu-MH}
        \end{figure*}

        We plot in Fig.~\ref{fig:Abu-MH} the abundance ratios [X/Fe] versus [Fe/H] for the OCCASO (circles) and OCCASO+ (triangles) samples, color-coded by age. To make the interpretation simpler, the different elements are sorted by their nucleosynthetic group.

        All $\alpha$ elements (Mg, Si, Ca, and Ti) show a slightly decreasing trend with metallicity, which is clearer for the OCCASO+ sample since it covers a larger [Fe/H] range.
        This trend, widely reported in the literature, is explained by the production of $\alpha$ elements mainly in core collapse supernovae (CCSs) from massive stars in short timescales in comparison with Fe, which is produced on longer timescales mostly by type Ia Supernovae (SNe Ia).
        The slopes of each $\alpha$ element can be different because of their different production chains \citep[e.g.,][]{Magrini2017}.

        The odd-z elements (Na and Al) seem to show a mild decreasing trend at subsolar metallicities up to [Fe/H] $\sim -0.2$.
        Even though there are large uncertainties involved, the Na abundance seems to increase at super-solar metallicities, while Al trend remains mostly flat.
        These elements are produced by massive and asymptotic giant branch (AGB) stars, but Na is also synthesized in red giants, and mixing effects bring the Na to the surface. The process is more important in massive giants, and thus Na can appear enhanced in young OCs (see Sect.~\ref{sec:age_gradient}).

        The Fe-peak elements (Sc, V, Cr, Mn, Co, Ni, Cu, and Zn) are likely to be produced by different nucleosynthesis processes. On the one hand, Sc, V, Cr, Mn, Co, and Ni are thought to be produced by the same processes as Fe \citep{kobayashi2020}, with a trend generally flat. Nevertheless, Sc and Co show mild decreasing trends at low metallicities, similarly to Al. On the other hand, the nucleosynthesis of the elements Cu and Zn is under debate \citep{Bisterzo2005,Romano2007,prantzos2018,kobayashi2020}. The different processes suggested relate their formation to massive stars. Cu exhibits a larger scatter, which may be attributed to higher uncertainties (computed as the abundance standard deviation), making it difficult to extract further conclusions. However, Zn appears to show a decreasing trend with Fe abundance, that could be compatible with its formation in massive stars.

        Neutron capture elements (Sr, Y, Zr, Ba, Ce, and Nd) show larger scatters than the others, mostly due to their age dependence (see Sect.~\ref{sec:age_gradient}), and also because of their larger uncertainty.
        These elements can be produced by slow (s) or fast (r) processes of neutron capture.
        They are defined by whether the capture timescale is longer or shorter than $\beta$ decay, and occur at different astrophysical sites.
        The s-process occurs mainly in AGB stars \citep[e.g.,][]{Gallino1998}, while the origin of the r-process is still under debate \citep[e.g.,][]{Kajino2019}.
        All neutron-capture elements studied in this work are produced by both processes, with different relative contributions.
        They all show similar general behavior with a dependence on age, being older OCs more depleted.
        Y, Zr, Ba, and Ce show a slight increasing trend with [Fe/H], reaching their maximum at [Fe/H]$\sim$0\,dex to decrease again more abruptly at higher [Fe/H] abundances.
        This last decrease is explained since as [Fe/H] increases, the ratio of neutrons to Fe in AGB star decreases. As a result, there is a smaller proportion of s-process elements being produced \citep{Gallino2006,Cristallo2009,Karakas2014}.
        Nd is the element which has the highest percentage production by r-process, 38\% according to \citet{Prantzos2020}. 
        It shows a steeper and more continuous decreasing trend with [Fe/H] compared to the other elements, which can be a consequence of its higher production by r-process.
        
        The obtained abundances patterns are generally compatible with those previously reported in the literature for thin-disk stars \citep[e.g.,][]{Adibekyan2012,Delgado-Mena2017,Delgado-Mena2019,Mikolaitis2019,Tau2021}.
        However, the increase of [Mn/Fe] with [Fe/H] reported in the literature \citep[e.g.,][]{Adibekyan2012,Mikolaitis2019} is not clearly seen in our case.
        We also find differences in the neutron capture maximum abundance compared to what is stated in the literature.
        While \citet{Delgado-Mena2017} finds it at [Fe/H] $\sim0$\,dex for the five elements similarly to us, \citet{Tau2021} finds the maximum around -0.2\,dex for Ba and Ce.
        In general, for one-zone Galactic chemical evolution models, the maximum is predicted at lower Fe abundances than seen in the observations \citep[e.g.,][]{Bisterzo2017, prantzos2018, kobayashi2020}.

\subsection{Abundance dependence on age}
    \label{sec:age_gradient}

    \begin{figure*}
    \centering
    \includegraphics[width=18cm]{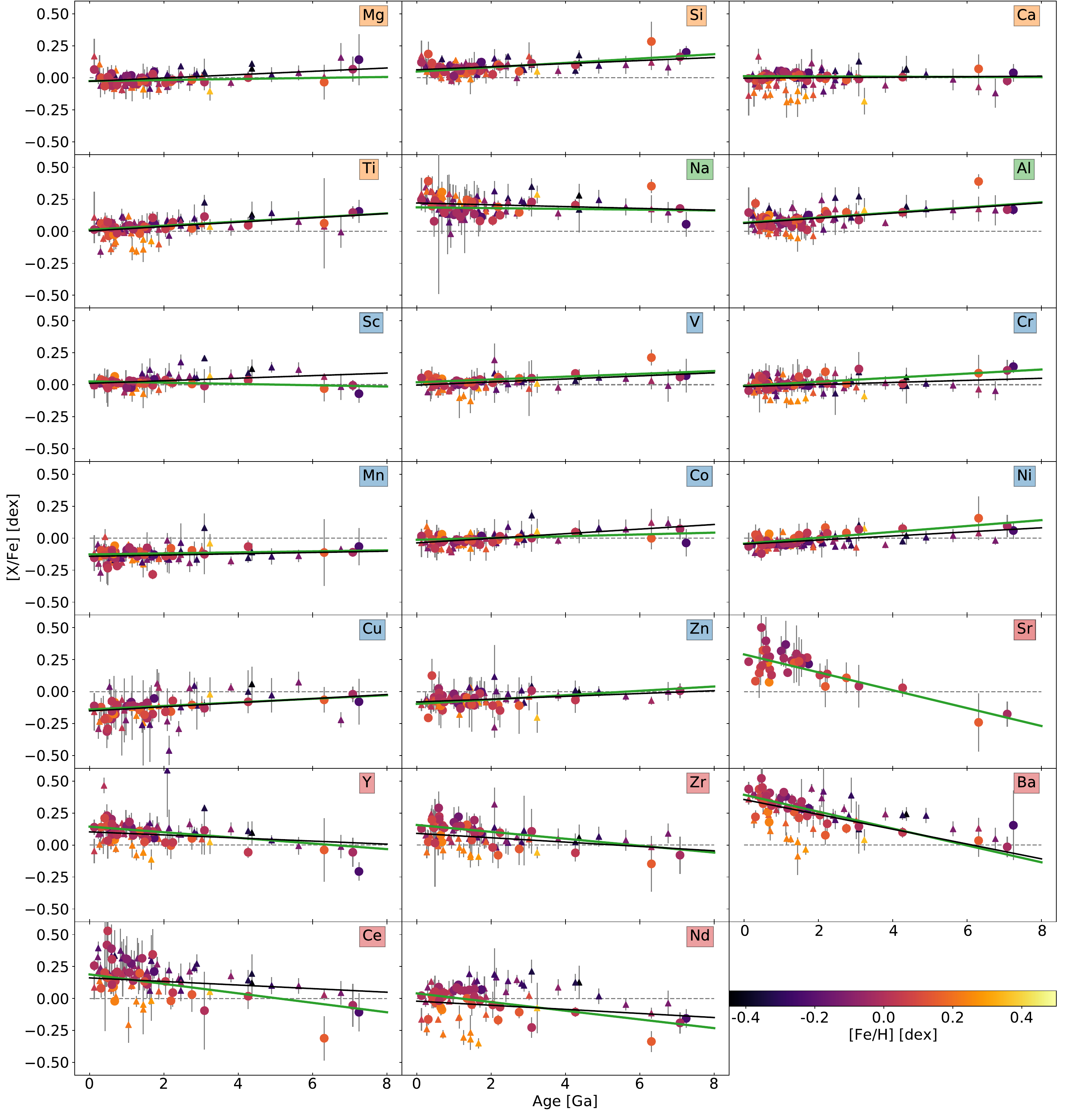}
    \caption{[X/Fe] ratios vs age, color-coded as a function of [Fe/H]. The symbols and panels are equivalent to  Fig. \ref{fig:Abu-MH}. The best fits for OCCASO (green) and OCCASO+ (black), respectively, are plotted.}
    \label{fig:age_g}
    \end{figure*}

    We plot the dependence of the different elements with age in Fig.~\ref{fig:age_g} where OCs are color-coded by their [Fe/H] abundance.
    For each element, we perform a linear fit to quantify its age dependence.
    These fits were performed by the same method as in \citet{Anders2017}, using a maximum likelihood algorithm as first guess, and computing a Markov-Chain Monte-Carlo with the python package \emph{emcee} \citep{Foreman2013}.
    The results of all fits are listed in the Table~\ref{tab:alpha-abu-age_OCC}.
    The best fits for both OCCASO and OCCASO+ samples are overplotted in Fig.~\ref{fig:age_g}, with green and black lines, respectively.

    From Mg to Co, we see very small dispersions, highlighting the small variability of these abundance ratios in the age and metallicity range covered by OCs of the thin disk.
    Nevertheless, due to the high-precision of our abundances (which is maximized thanks to the number of stars per cluster) we are able to see some mild trends.
    
    In particular, we obtain positive slopes for the $\alpha$ elements Si and Mg, which has been reported in the literature using high-precision samples of field stars \citep[e.g.,][]{Delgado-Mena2019}.
    Furthermore, we noticed enhanced levels of these elements in several young inner disk OCs aged between 0.1 to 0.7 Ga. Specifically, increased levels of Mg are observed in NGC 6067, NGC 6259, and UBC 3 OCs, while higher Si levels are found in NGC 6067 and NGC 6705 OCs.
    In the literature, the cluster NGC\,6705 was found to be $\alpha$ enhanced \citep{Magrini2014,casamiquela2018}, and was considered as a peculiar OC. 
    To our knowledge, this is the first time that NGC\,6705 is reported to belong to a group of $\alpha$ enhanced clusters in the inner disk.  
    
    The two odd-Z elements Na and Al exhibit different behaviors.
    Al shows an overall increasing trend consistent with the results of \citet{Delgado-Mena2019}. 
    Na is mostly flat, though we remark a mild enhancement at young ages until $\sim$1.8\,Ga, and a plateau for older ages.
    The Na-enhancement can be explained by the fact that the atmospheres of massive red giant stars (1.5-2\,M$_\odot$) are polluted after the first dredge-up because of deep mixing \citep{Smiljanic2016,Lagarde2012}.
    As a consequence, OCs younger than $\sim$1.2--2.5\,Ga can appear enhanced in Na \citep[see also][]{Casamiquela2020}.
    
    Fe-peak elements show some variability in their trends as a function of age, with part of them being positive in both fits, in particular V, Co, Ni, Cu, and Zn.
    \citet{Mikolaitis2019} have also found clear positive trends for Co and Ni in agreement with us, but they find a negative trend for Cu and Mn, in contrast with our results.
    On the other hand, \citet{Delgado-Mena2019} reports similar trends as ours in Cu and Zn.

    Neutron capture elements are known to have high dependencies with age.
    We find steep decreasing trends for all the elements comparable to what has been found in the literature, in particular, \citet{Casamiquela2021} which used a sample of 47 OCs that included most of the OCCASO sample.
    The trends are also in agreement with those found by \citet{Viscasillas2022} for GES OCs, and those reported by \citet{Spina2016} and \citet{Delgado-Mena2019} for field stars. 
    The enhancement of s-process elements in young stars is tentatively explained in chemical evolution models by assuming enhanced AGB yields of low-mass stars \citep[e.g.,][]{D'Orazi2009, Cristallo2015}.
    These low-mass stars may take several Ga to deliver their chemical products into the interstellar medium, leading to a delay in the enrichment of s-process elements, and therefore, to a strong [X/Fe]-age dependence. What is more, the age dependence does not have to be monotonic. In the trend of Ba and possibly Zr there is a hint of a flattening at ages older than $\sim$2.5\,Ga (see Fig.~\ref{fig:age_g}).
    A similar flattening is reported for Ce by \citet{Sales-Silva2022} from the APOGEE DR17 OC sample, but at 4\,Ga. 
    This kind of non-monotonic trend could be caused by a change in the enrichment rate of the interstellar medium by AGB stars.
    However, we also recall that the computation of Ba abundances can significantly be affected by activity in young stars \citep{Reddy2017,Spina2020}, which can alter the Ba trend with age. 

\begin{figure}
    \centering
    \includegraphics[width=\columnwidth]{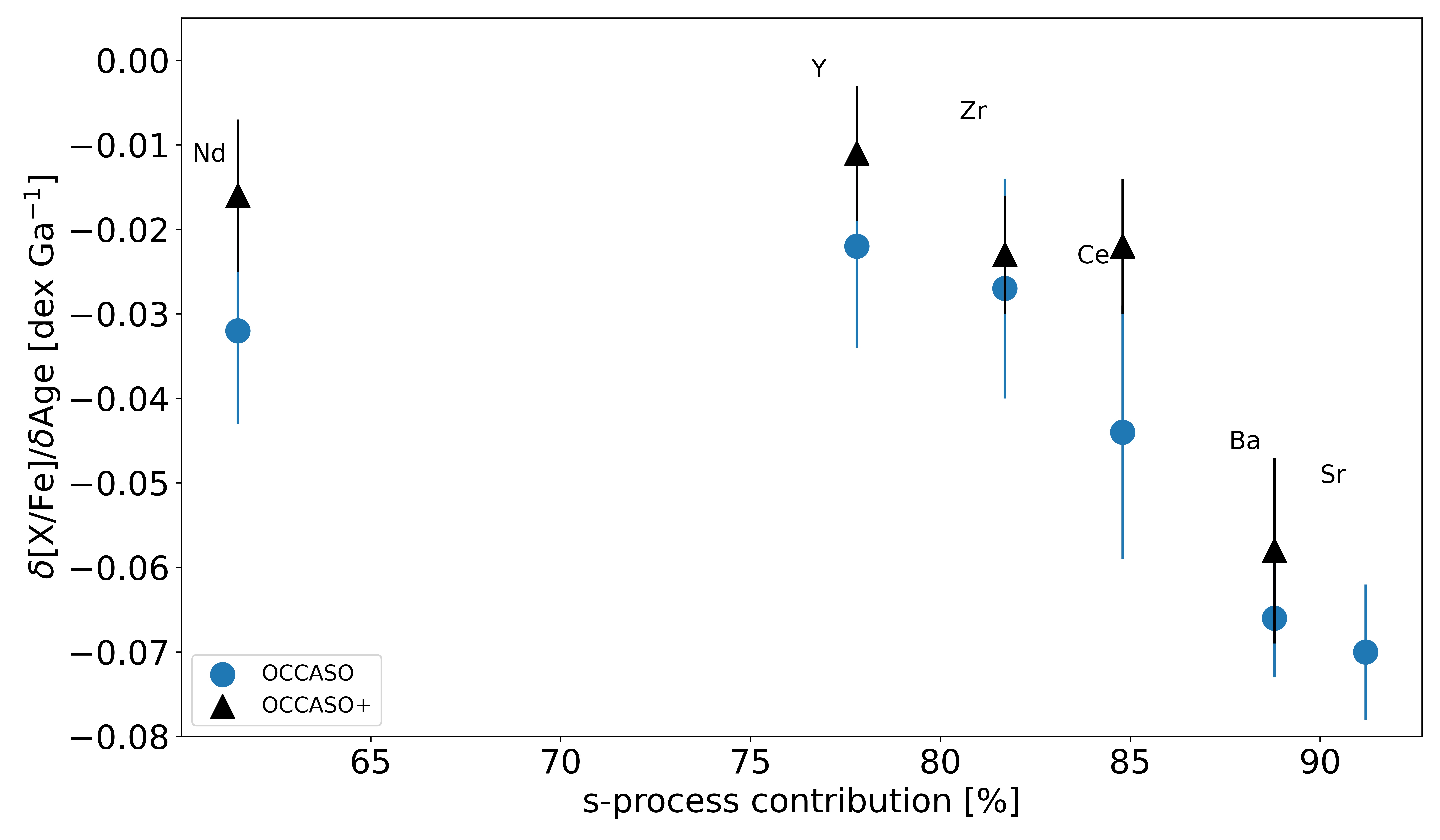}
    \caption{Slopes of the [X/Fe]–age relations of the neutron-capture elements as a function of their s-process contribution percentages computed by \citet{Prantzos2020}.} 
    \label{fig:s_var_sperc}
    \end{figure}

    
    Each element has a different trend according to the ratio at which it is formed by s- and r- processes, as shown in Fig.~\ref{fig:s_var_sperc}.
    The slopes get steeper with increasing contribution of s-process.
    This dependence has also been found for solar twins \citep{Spina2016} and field stars \citep{Delgado-Mena2019}.
    It can be explained by the different production timescales of the two processes.
    Unlike the s-process, r-process is expected to occur on short timescales, shorter than those of Fe production.
    Therefore, the more an element is produced per s-process, the steeper its [X/Fe]-age dependence would be.
    Several authors calculate the s-process contribution \citep[e.g.,][]{Bisterzo2014,Prantzos2020}, giving diferent values. The element for which their results differ the most is Sr, being 68.9\% at \citet{Bisterzo2014} and 91.2\% at \citet{Prantzos2020}. Our results agree better with the value of \citet{Prantzos2020}.


    \begin{figure}[hbt!]
    \centering
    \includegraphics[width=\columnwidth]{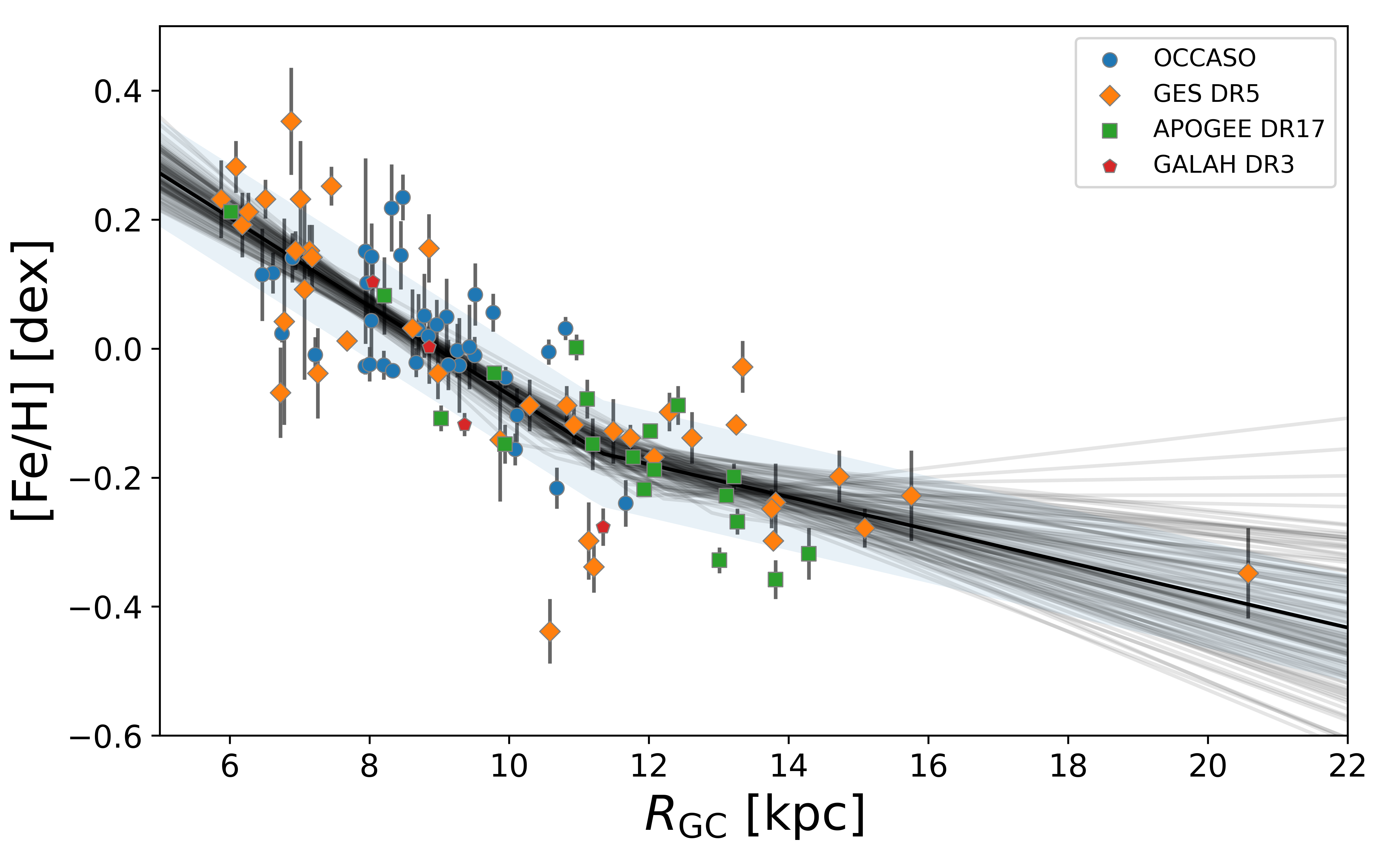}
    \caption{[Fe/H] versus Galactocentric radius for OCCASO+ sample. The different surveys are color-coded as in Fig.~\ref{fig:XY}. The grey vertical lines represent the uncertainties of abundances, and the black line represents our best fit.}
    \label{fig:0.18}
    \end{figure}

\subsection{Radial trends}
    \label{sect:radial_trends}

        In this section, we study the dependencies on the $R_{\rm GC}$ of [Fe/H] and abundance ratio to Fe of the rest of the elements.    
        The radial distribution of [Fe/H] is one of the most widely used tracers in Galactic archaeological studies.
        The general consensus is that there is a steeper decreasing gradient in the inner disk and a plateau in the outer regions, the so-called knee shape.
        In this section, we investigate the radial trends on [Fe/H] vs $R_{\rm GC}$ obtained with the OCCASO and OCCASO+ sample. We use the fitting procedure described in Sect.~\ref{sec:age_gradient}, but in the case of OCCASO+ we model the fit with two lines, adjusting at the same time the knee position. The OCCASO sample contains only OCs inside the knee. 
        %
        Table~\ref{tab:FeH} contains the slopes obtained with both samples compared to recent literature studies. 
        The knee position derived with OCCASO+ is 11.3 $\pm$ 0.8 kpc which confirms the position found in the latest studies in te literature.   
        The results obtained for both OCCASO and OCCASO+ samples inside the knee are compatible with previous determinations in the literature.
        Outside the knee, we recover a flatter trend for OCCASO+ in comparison with GES23, and more similar to \citet{Spina2022}, \citet{myers2022} and \citet{carrera2019}.
        We remark that this result is independent of the inclusion of Berkeley~29, the furthest cluster in the sample.

    \begin{table*}[h!]
    \setlength{\tabcolsep}{1mm}
    \begin{center}
    \caption{Comparison of [Fe/H] radial gradient with the literature in the region inside and outside the knee radius and globally, indicating in each case the number of OCs studied and the knee position. }
    \begin{tabular}{lcccccccc}
    \hline
    Reference & Inside the knee radius & $N$ & Outside the knee radius & $N$ & Global & $N$ & Knee\\
     & [dex kpc\textsuperscript{-1}] &  & [dex kpc\textsuperscript{-1}] &  & [dex kpc\textsuperscript{-1}] & & [kpc] \\
    \hline
    This work OCCASO & -0.059$\pm$0.017 & 36 & -- & -- & -- & -- & -- \\
    This work OCCASO+ & -0.069$\pm$0.008 & 71 & -0.025$\pm$0.011 & 28 & -0.062$\pm$0.007 & 99 & 11.3 $\pm$ 0.8\\
    Paper III & -0.056$\pm$0.011 & 18 & -- & -- & -- & -- & -- \\ 
    \cite{carrera2019} & -0.077$\pm$0.007 & -- & -0.018$\pm$0.009 & -- & -0.077$\pm$0.007 & 90 & 11\\
    \cite{Donor2020}  & -0.068$\pm$0.004 & 68 & -0.009$\pm$0.011 & 3 & -- & 71 & 13.9\\
    \cite{Zhang2021}  & -0.066$\pm$0.005 & 157 & -0.032$\pm$0.007 & 4 & -- & 161 & 14\\
    \cite{myers2022} & -0.073$\pm$0.002 & 51 & -0.032$\pm$0.002 & 34 &  -0.055 $\pm$0.001 & 85 & 11.5\\ 
    GES23 & -0.081$\pm$0.008 & 42 & -0.044$\pm$0.014 & 20 & -0.054$\pm$0.004 & 62 & 11.2 \\ 
    \citet{Spina2022}  & -0.064$\pm$0.007 & -- & -0.019$\pm$0.008  & -- & -- & -- & 12.1 $\pm$ 1.1\\
    \cite{Netopil2022} & -0.063$\pm$0.004 & 116 & -- & -- & -0.058$\pm$0.005 & 136 & 12\\
    \citet{GC_recioblanco2022} & -0.054$\pm$0.008 & 503 & -- & -- & -- & -- & -- \\

    \hline
    \end{tabular}
    \label{tab:FeH}
    \end{center}
    \end{table*}

        \begin{figure*}
        \centering
        \includegraphics[width=18.5cm]{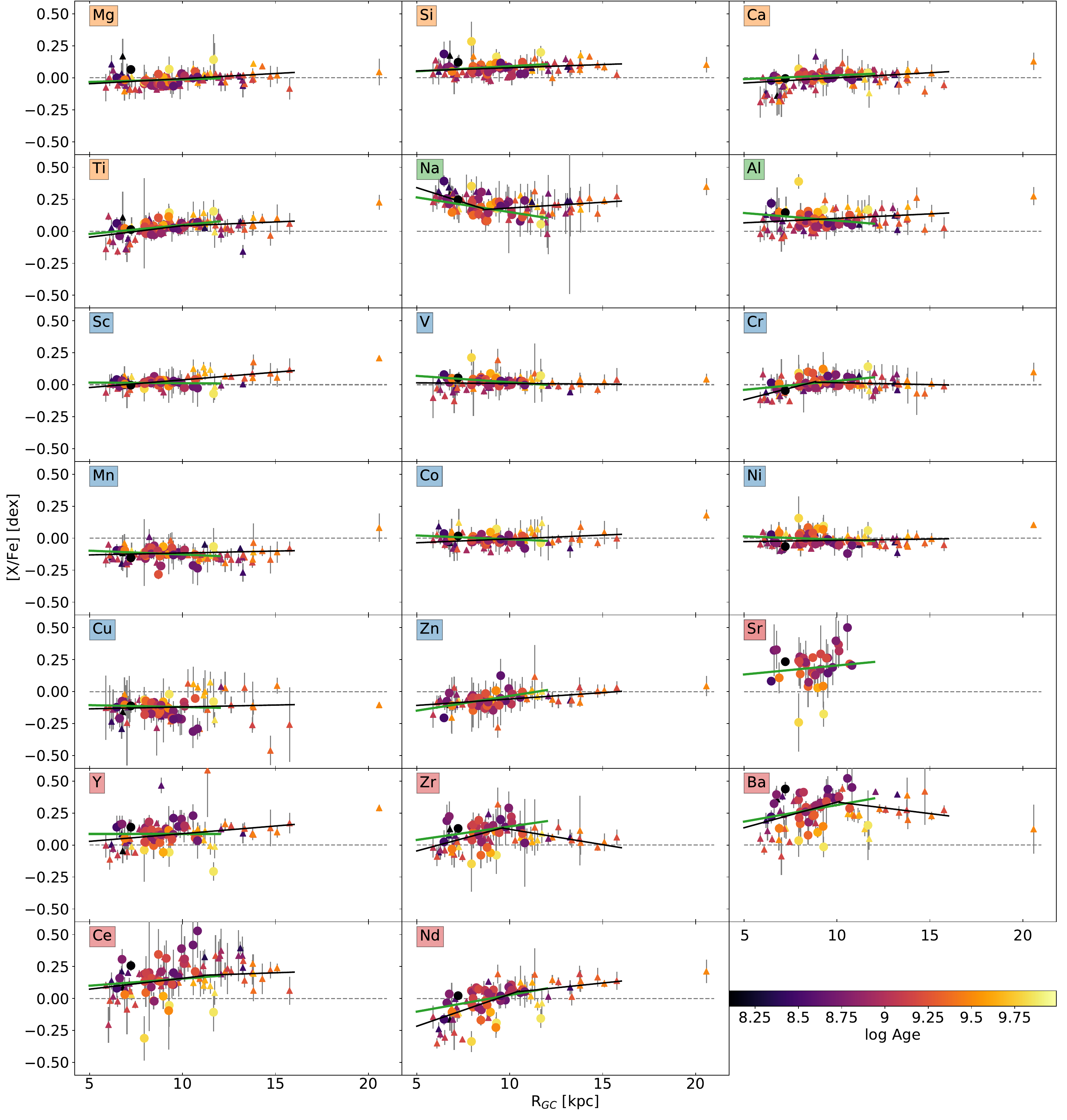}
        \caption{Abundance [X/Fe] ratios as function of $R_{\rm GC}$ for the OCCASO (circles) and OCCASO+ (triangles) samples color-coded with age of the OCs. Fitted lines for both samples are shown in green and black, respectively.}
        \label{fig:Rgc}
        \end{figure*}

     \label{alfa}

    In Fig.~\ref{fig:Rgc}, we show the dependence of the different studied elements [X/Fe] with $R_{\rm GC}$ for the OCCASO (circles) and OCCASO+ (triangles) samples and their corresponding linear fits (green and black).
    The OCCASO+ sample has been fitted with two lines as in the case of the [Fe/H] trends. We have limited the sample to 16 kpc, since at larger distances we only have one cluster in the sample, and therefore this region is not sampled correctly. The code used allows us to determine if a knee shape is found or not. 
    For more than half of the elements, there is no knee shape, and we fit the whole sample with a single line. We find a knee shape for the elements Ti, Na, Cr, Y, Zr, Ba, Ce, and Nd. The results of the analysis are listed in Table \ref{tab:OPLUS_all}.

    
        The $\alpha$ elements show mild positive gradients, more evident for Ca Mg, and Ti.
        This is in agreement with the models of the inside-out formation of the Galactic disk.
        The differences among the elements, as already mentioned, are due to their different detailed production processes.
        
        The odd-z elements show different behaviors. Al has a flat relation whereas Na shows a decrease up to 10\,kpc showing a positive trend from there off. As already discussed, the enrichment of the Na atmospheric abundances along the red giant phase makes difficult to reach further conclusions. 

        Analogous to the discussion in Sect.~\ref{sect:metallicity}, we expected Fe-peak elements produced by the same processes as Fe to have flat radial gradients. We have observed this for V, Mn, Co, and Ni inside the knee. However, Cr shows a positive gradient inside the knee, which seems to decrease again outside it. On the contrary, we note that Sc shows a flat trend in the OCCASO sample and a positive trend in the OCCASO+ sample. Additionally, elements mainly produced in massive stars should exhibit positive radial trends, as some $\alpha$ elements do. It could also be the case for Zn, showing a positive trend in the whole studied range.

        In the case of the neutron-capture elements, we see again larger dispersions at a given $R_{\rm GC}$ due to their age dependence and larger uncertainties.
        They tend to show positive gradients, possibly due to the dependence of s-process production with metallicity.
        However, there are significant differences among them in agreement with the fact that each element is produced in different proportions by r- and s-processes (as discussed in Sect.~\ref{sec:age_gradient}).
        We find flatter trends for Sr, Y, Zr, and Ce, and steeper trends for Ba and Nd.
        In general, the trends are flatter outside the knee for all the elements. That is due to the dependence of s-process on [Fe/H] previously discussed. 
        In the case of Sr we do not know if it happens, as we have no observations outside the knee.

\begin{figure*}
        \centering
        \includegraphics[width=18.5cm]{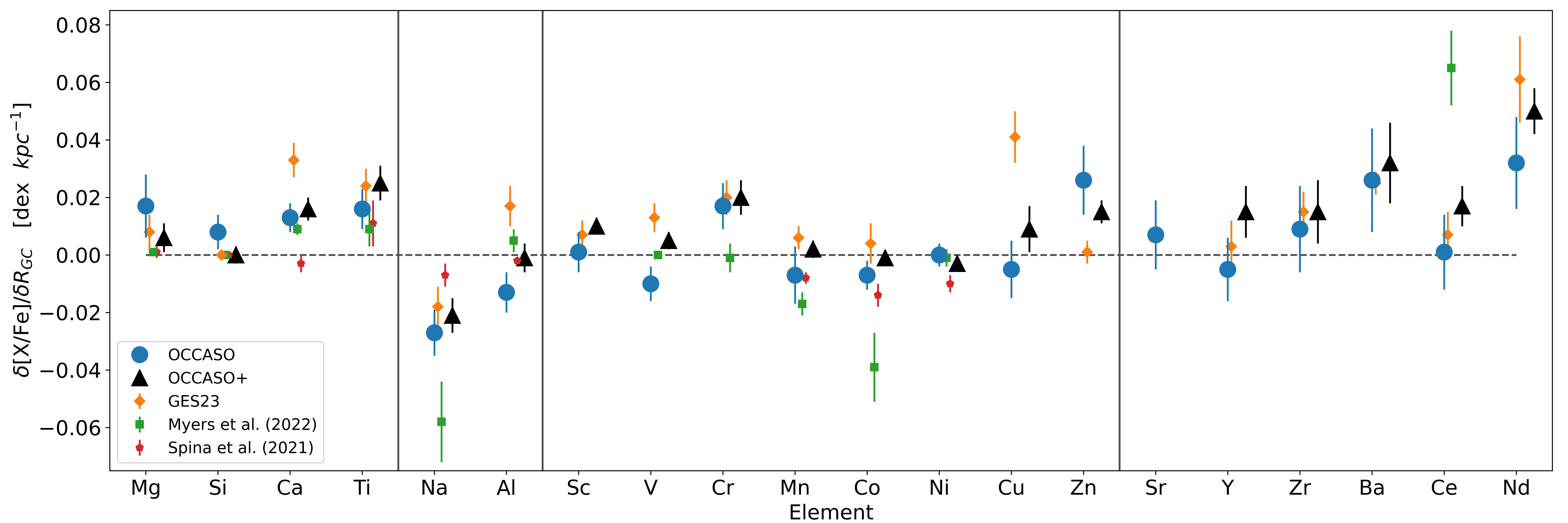}
        \caption{Comparison of [X/Fe] radial gradients with the literature.}
        \label{fig:trends}
        \end{figure*}

    In order to compare our results with the latest studies, we analyzed the samples published by GES23, \citet{myers2022}, \citet{spina2021}, and OCCASO+ restricting the analysis to the same $R_{\rm GC}$ range that in OCCASO sample.
    The comparison is shown in Fig.~\ref{fig:trends}.
    Overall, we find good agreement between the different samples taking into account the uncertainties, although there are a few exceptions.
    We highlight that we find large differences between \citet{spina2021} and our sample in Ca and Na, and between \citet{myers2022} and our sample in Na, Co, and Ce.
    The different slope of Co may be due to the dispersion and high uncertainties of the APOGEE abundances in this element.
    We also find remarkable differences of Ca, Al, Cu, V, and Zn gradients for GES23 compared with the other samples.
    For some of these elements (Ca, Ti, Al, and V) the difference in the gradient is due to the fact that in the inner disk ($R_{\rm GC}$ $\leq$ 8\,kpc) some OCs in GES23 tend to have lower abundances compared to other surveys (see Fig.~\ref{fig:AnexSComp}).
    Some of these OCs in GES23 are in common with \citet{myers2022} (Trumpler~20 and Ruprecht~134), which does not confirm their low abundances.
    Indeed, GES23 abundances of those OCs are $\sim$0.07 dex lower than \citet{myers2022} values.
    Given the way in which the OCCASO+ sample is constructed, we believe that its results are the most robust and reliable.

\subsection{Temporal evolution of Galactocentric radial gradients}
\label{sect:temp_evo_rad}

      \begin{table*}[h!]
    \setlength{\tabcolsep}{1mm}
    \begin{center}
    \caption{Change with age of the [Fe/H] radial gradients studied through MCMC at Sect.~\ref{sect:radial_trends}.}
    \begin{tabular}{lccccccc}
    \hline 

Sample&Age bin&b1&m1&m2&Knee&N&Spearman\\
&[Ga]&[dex]&[dex kpc$^{-1}$]& [dex kpc$^{-1}$]&[kpc]& \\
    \hline 
OCCASO&0.1-1&0.365 $\pm$ 0.211&-0.032 $\pm$ 0.022&&&15&-0.30\\
OCCASO&1-2&0.457 $\pm$ 0.084&-0.079 $\pm$ 0.015&&&11&-0.44\\
OCCASO&2-3&0.347 $\pm$ 0.285&-0.054 $\pm$ 0.037&&&4&-0.47\\
OCCASO&3-7.3&0.828 $\pm$ 0.553&-0.099 $\pm$ 0.023&&&5&-0.95\\
    \hline
OCCASO+&0.1-1&0.452 $\pm$ 0.062&-0.049 $\pm$ 0.006&&&39&-0.75\\
OCCASO+&1-2&0.795 $\pm$ 0.114&-0.09 $\pm$ 0.014&-0.019 $\pm$ 0.018&10.1 $\pm$ 0.8&29&-0.87\\
OCCASO+&2-3&0.651 $\pm$ 0.275&-0.075 $\pm$ 0.033&-0.042 $\pm$ 0.017&9.6 $\pm$ 2.1&14&-0.85\\
OCCASO+&3-8&0.996 $\pm$ 0.305&-0.113 $\pm$ 0.034&-0.012 $\pm$ 0.018&10.9 $\pm$ 1.1&15&-0.80\\
    \hline
\citet{myers2022}&0.1-1&0.395 $\pm$ 0.087&-0.051 $\pm$ 0.008&&&39&-0.72\\
\citet{myers2022}&1-2&0.673 $\pm$ 0.135&-0.08 $\pm$ 0.016&-0.041 $\pm$ 0.01&10.1 $\pm$ 1.0&20&-0.90\\
\citet{myers2022}&2-3&0.687 $\pm$ 0.177&-0.085 $\pm$ 0.017&-0.014 $\pm$ 0.026&11.9 $\pm$ 0.9&11&-0.77\\
\citet{myers2022}&3-8&1.455 $\pm$ 0.329&-0.161 $\pm$ 0.035&-0.025 $\pm$ 0.017&10.8 $\pm$ 0.6&12&-0.70\\
    \hline
\citet{spina2021}&0.1-1&0.558 $\pm$ 0.062&-0.065 $\pm$ 0.006&&&57&-0.72\\
\citet{spina2021}&1-2&0.566 $\pm$ 0.187&-0.068 $\pm$ 0.019&-0.037 $\pm$ 0.047&11.5 $\pm$ 1.3&22&-0.86\\
\citet{spina2021}&2-3&0.83 $\pm$ 0.234&-0.097 $\pm$ 0.024&-0.019 $\pm$ 0.028&11.9 $\pm$ 1.1&10&-0.93\\
\citet{spina2021}&3-7.3&1.307 $\pm$ 0.295&-0.147 $\pm$ 0.032&-0.02 $\pm$ 0.018&11.1 $\pm$ 0.8&16&-0.82\\
    \hline
GES23&0.1-1&0.356 $\pm$ 0.057&-0.043 $\pm$ 0.006&&&23&-0.82\\
GES23&1-2&0.811 $\pm$ 0.152&-0.09 $\pm$ 0.019&-0.02 $\pm$ 0.026&10.8 $\pm$ 1.0&17&-0.85\\
GES23&3-7.3&1.025 $\pm$ 0.34&-0.118 $\pm$ 0.038&-0.014 $\pm$ 0.017&10.7 $\pm$ 1.1&12&-0.75\\
    \hline
\gaia DR3&0.1-1&0.384 $\pm$ 0.069&-0.043 $\pm$ 0.005&&&303&-0.43\\
\gaia DR3&1-2&0.588 $\pm$ 0.196&-0.065 $\pm$ 0.01&&&48&-0.55\\
\gaia DR3&2-3&0.933 $\pm$ 0.392&-0.079 $\pm$ 0.025&&&25&-0.41\\
\gaia DR3&3-7.3&0.354 $\pm$ 0.26&-0.105 $\pm$ 0.021&&&13&-0.54\\

    \hline
    \end{tabular}
            \tablefoot{The columns b1 and m1 are the y-intercept and slope of the first line fitted, respectively. The column m2 is the slope of the second line. The position of the knee, the number of OCs and Spearman correlation coefficient are listed in the last columns. We analyzed the OCCASO and OCCASO+ samples and reanalyzed the others. The OCCASO and \gaia DR3 samples do not have clusters beyond the knee, so only the first line is fitted. In the first bin of all the samples, we did not find the knee shape.}
    \label{tab:AZ}
    \end{center}
    \end{table*}

\begin{figure}[hbt!]
    \centering
    \includegraphics[width=\columnwidth]{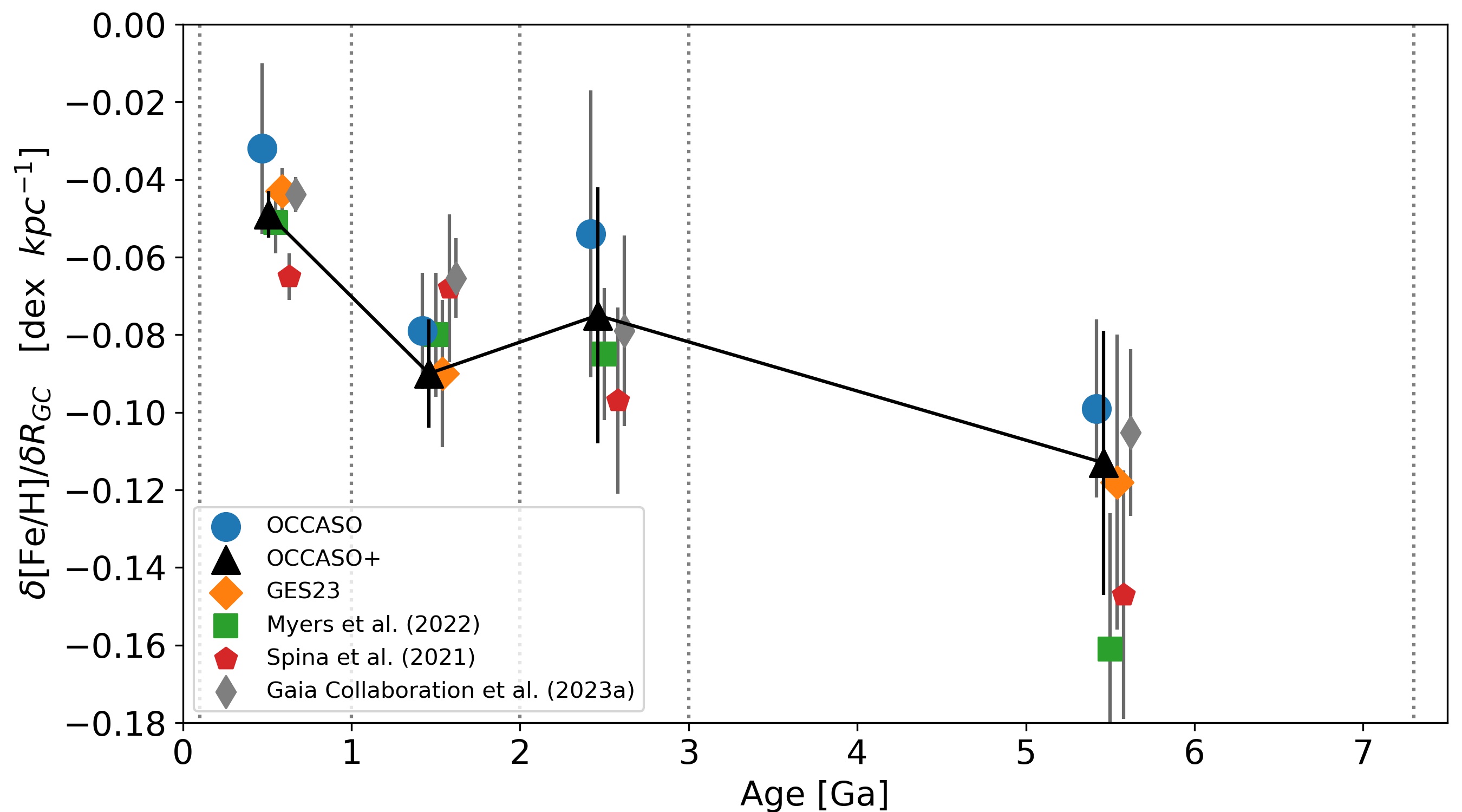}
    \caption{Evolution of the [Fe/H] radial gradient with age for the different samples analyzed. The gradients are those measured inside the knee. The age position of each sample is slightly changed for clarity of the plot.}
    \label{fig:FeA1}
    \end{figure}

        In order to investigate the change of the [Fe/H] radial gradient with time, we consider OCs with $R_{\rm GC}$ < 16 kpc for the reasons given above, and we divide the samples used in Sect.~\ref{sect:radial_trends} into four age bins: 0.1-1, 1-2, 2-3, >3\,Ga, respectively. We maintain the same analysis methodology, providing the slopes of the two fitted lines and the knee position for each age bin (listed in Table~\ref{tab:AZ}). 
        The OCCASO+ sample shows the absence of knee in the youngest bin, and its presence in older clusters (see Fig.~\ref{fig:separated_age_bins}). 
        By studying the slope inside the knee, the OCCASO+ sample shows a steepening of the trend with age that seems to be non-linear (Fig.~\ref{fig:FeA1}), since the second age bin has a steeper gradient than expected considering the other age bins. We have tested that this non-linear dependence still appears when we change the way we partition the clusters into age bins. The slope outside the knee remains constant, taking into account errors. By re-analyzing the other samples, we obtain values compatible with OCCASO+, which is a confirmation of the dependence we find. This is an effect that does not appear in field stars, as has been found in the studies such as \citet{casagrande2011,Anders2017, Anders2023}. They show a steepening of the gradient between 1 and 2 Ga to become progressively flatter at older ages, contrary to OCs. The flattening of the gradient is an expected effect of radial migration.       

        There are two hypotheses that attempt to explain the change in the gradient with age in OCs: 
        \citet{Anders2017} proposed that the change of trend is produced by the radial migration process coupled with a selective bias that the Galaxy exerts on OCs. Only outward migrating clusters survive because the Galactic potential becomes less destructive as the distance from the centre increases. Inward migrating clusters are quickly disrupted. 
        Otherwise, GES23 show in their Fig. 5 that young clusters (<1Ga) in the inner disk have lower Fe abundances than older OCs, and propose that a considerable infall of gas with low metallicity has produced the last episode of star formation. In the OCCASO sample, we do not clearly see this age separation. In fact, there are two young OCs in the inner disk that are metal-rich: NGC~2632 and NGC~6997 in contradiction with this hypothesis (Fig.~\ref{fig:separated_age_bins}). This should be studied further, with larger samples in the inner disk.  
        
        Another issue for which there is still no satisfactory explanation is the presence of the plateau in the radial gradient. This feature is seen in OCs, and recently in cepheids \citep{daSilva2023}.
        Similar hypotheses to the ones above have been suggested to explain it: \citet{magrini2009} suggested that merger events or infall from the halo, affecting large radii and providing pre-enriched material, could explain the plateau. 
        Another possible explanation is that the plateau is produced by radial migration of OCs towards the outer disk. We do not know the event or mechanism that could have produced it, though. 
        The OCCASO+ sample shows no plateau in the youngest age bin (<1Ga), with OCs extended up to 14 kpc. We have ruled out that we have a radial bias in the sample, as in the \citet{cantat-gaudin2020} catalog this is the maximum distance at which clusters appear in this age range. As the knee formation event does not affect young clusters, it must be a process that occurred more than 1 Ga ago. The presence of OCs older than 1 Ga at more than 14 kpc, together with the appearance of the knee from this epoch, supports this hypothesis. 
        The recent cepheid study by \citet{daSilva2023} shows signs of flattening in the outer disk. Given the youth of these stars, the presence of cepheids forming a plateau cannot be explained by radial migration processes. Further research is needed on both tracers in the outer disk.

\begin{figure*}
\centering
\includegraphics[width=18.5cm]{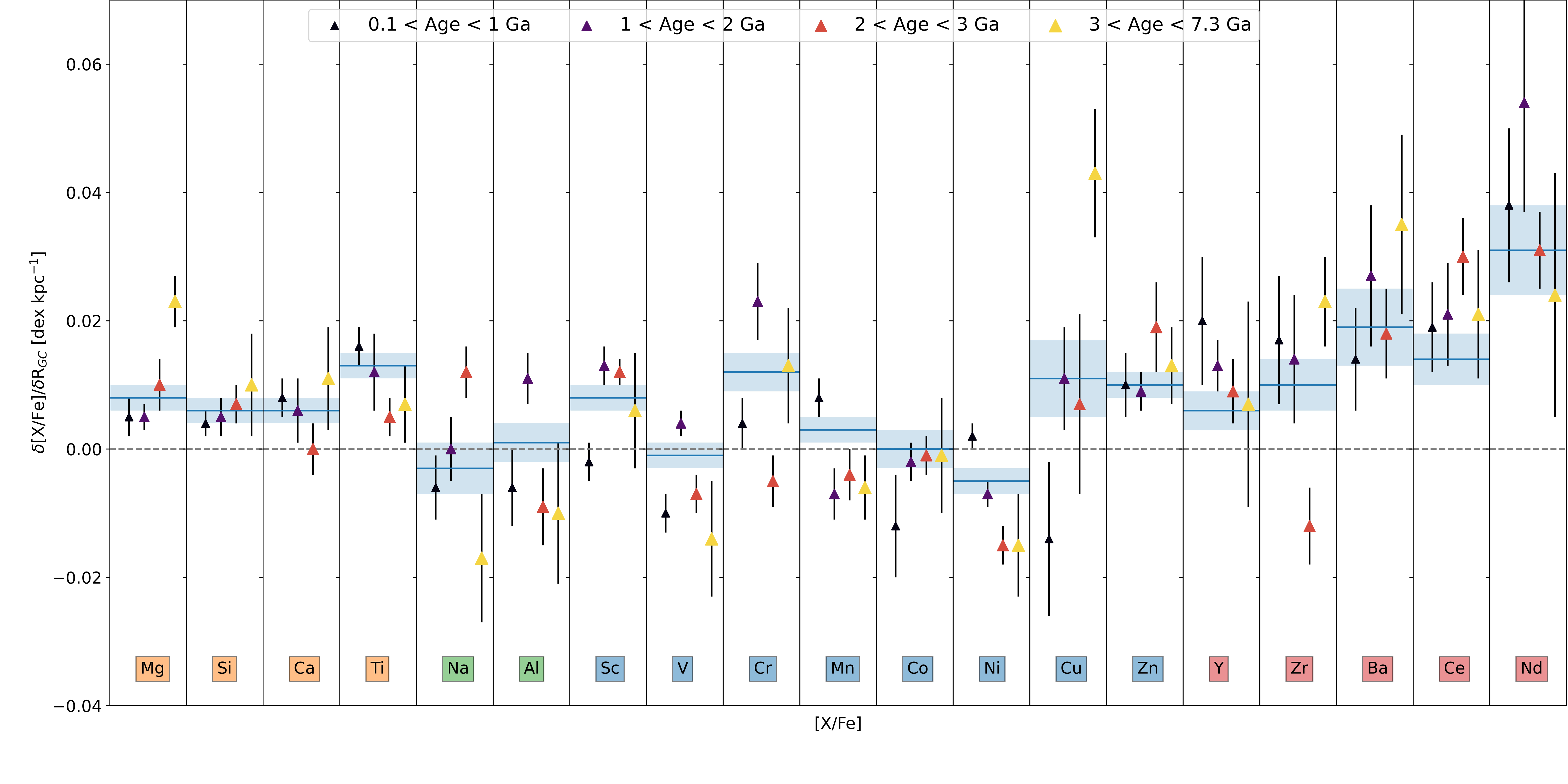}
\caption{Change of OCCASO+ [X/Fe] radial gradient in four age bins, depicted by the size and the color of the markers. In each panel, age is growing towards the right. Blue horizontal lines represent the radial gradient for the whole age range, and the shadow area shows its dispersion. The table with the trends is available under request.}
\label{fig:XFe_Rgc_Age}
\end{figure*}

    Figure~\ref{fig:XFe_Rgc_Age} shows the change of $\delta$[X/Fe]/$\delta R_{\rm GC}$ as a function of age of the OCCASO+ sample. Filled symbols represent the slopes colored by the age range, and blue lines represent the gradient for all the OCs independently of age. 
    In cases where we did not find a knee in the global analysis, we  analyzed each age bin using clusters up to 16 kpc. Otherwise, in the cases where we found a knee, we evaluated the slope of the first line of the fit (m1 in Table~\ref{tab:AZ}). 
    We recall that when separating the samples by age, the amount of OCs per bin is small, and therefore the differences might be dominated by small number statistics.
    In most of the cases, the values in each bin are around the global one, without a clear correlation with age. Therefore, for almost all elements, if there is gradient age dependence, it should be lower than our uncertainties. 
    If we take as a reference the bin with the highest error, we can establish that the upper limit on the trend change is $\sim$ 0.01 dex kpc$^{-1}$. This value is of the order of 6 times lower than the gradient change we find for [Fe/H] indicating the absence of change with age of $\delta$[X/Fe]/$\delta R_{\rm GC}$. Or, in other words, $\delta$[X/H]/$\delta R_{\rm GC}$ changes with age essentially as $\delta$[Fe/H]/$\delta R_{\rm GC}$.
    However, there are hints of trend change for some elements. [Mg/Fe] seems to show an increase in the gradient with age, while [Ti/Fe] and [Ni/Fe] appear to decrease. More OCs and a better distribution among the different age bins are needed to address this topic in more detail.

\subsection{Azimuthal gradient} 
\label{sect:az}

    \begin{figure*}
        \sidecaption
          \includegraphics[width=17cm/2]{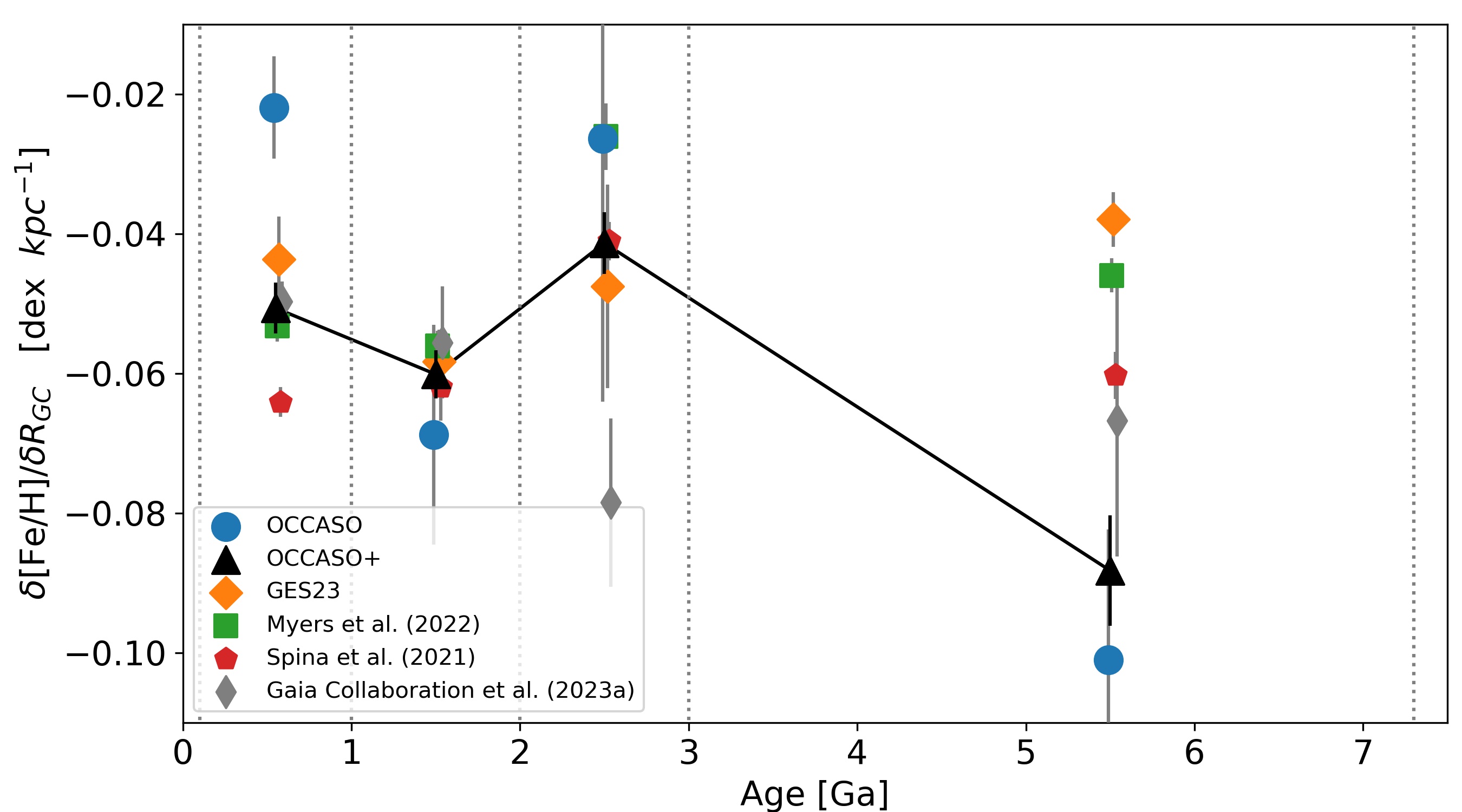}
          \hfill
          \includegraphics[width=17cm/2]{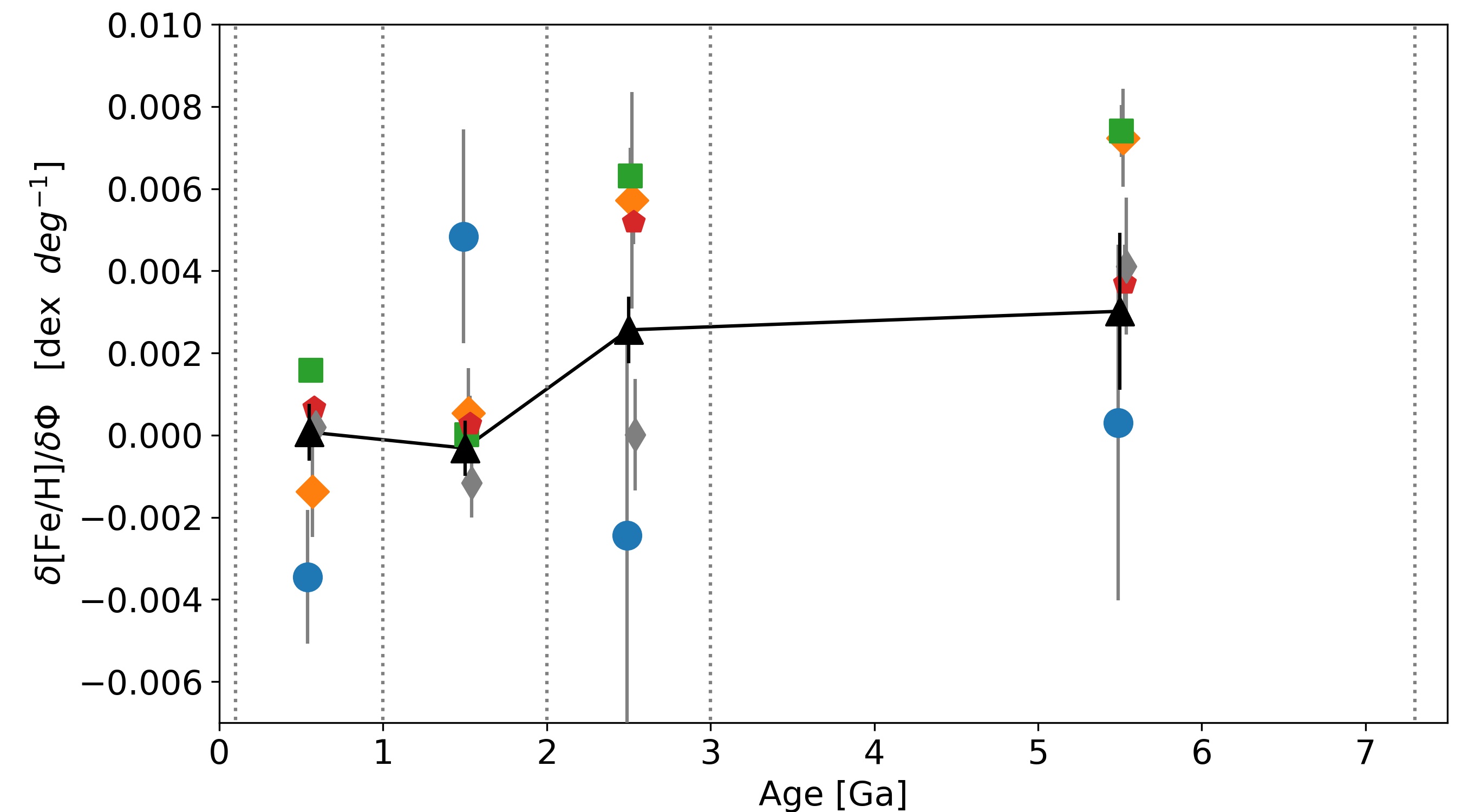}
        \caption{Evolution with age of the [Fe/H] radial gradient (left) analyzed at the same time as the azimuthal gradient (right).The age position of each sample is slightly changed for clarity of the plot.}
             \label{fig:az-rad_grad_az}
    \end{figure*}

    One interesting feature of the metallicity distribution of the OCs population as a function of Galactocentric radius is the large dispersion ($>$0.3\,dex) observed at any position (Figs.~\ref{fig:0.18} and \ref{fig:Rgc}). This dispersion cannot be explained by individual uncertainties and is usually attributed to the radial migration.
    During their lifetimes, stars, and OCs, can move from their birth Galactocentric radius due to the dynamical influence of non-axisymmetric structures in the Galaxy such as spiral arms \citep[e.g.,][]{Sellwood2002}, the bar \citep[e.g.,][]{Minchev2010} or minor satellites \citep[e.g.,][]{Quillen2009}. An alternative explanation is that the observed dispersion could be due to variations of the metal content with Galactic azimuth, as proposed by \citet{Friel2013}. Some evidence of abundance variations with the azimuth has been reported for Cepheids \citep{Luck2006} and young OCs \citep{Davies2009} but not for \mbox{H\,{\sc ii}} regions \citep{Arellano2020}. However, the lack of OC samples covering a wide azimuthal range has hampered the investigation of this hypothesis.

    We studied the same five samples analyzed in the previous section in order to check the existence of an azimuthal gradient. For this purpose, we fit simultaneously the radial and azimuthal gradients with OCs separated in the same age bins used in the previous section. To do that, we fit a plane using the \texttt{scikit-spatial}\footnote{https://scikit-spatial.readthedocs.io/en/stable/index.html} Python package. The uncertainties are constrained by generating 1000 possible values for each cluster assuming a Gaussian distribution centred on the mean abundance, with a $\sigma$ equal to its uncertainty. The resulting fits are listed in Table~\ref{tab:AZT} and shown in Fig.~\ref{fig:az-rad_grad_az} in the two projections ($R_{\rm GC}$, and azimuth $\Phi$). We do not represent bins containing fewer than ten OCs as we consider this to be insufficient statistics to conduct the study. 
    Compared to Fig.~\ref{fig:FeA1}  the evolution of the $R_{\rm GC}$ gradient (left panel) shows larger differences among samples, particularly in the oldest age bin, probably due to the additional dimension of the fit.
    In the case of OCCASO+ the obtained slopes are compatible with the results of Sect.~\ref{sect:temp_evo_rad}. 
    In the right panel of Fig.~\ref{fig:az-rad_grad_az} we show the evolution of the azimuthal gradient.
    The two youngest bins, in the OCCASO+ sample, do not show a dependence with azimuth, but there are hints of a positive trend in the oldest ones. This tendency also appears at GES23, \citet{myers2022}  and \citet{spina2021} samples. \citep{GC_recioblanco2022}, shows a flat gradient in the third age bin and a positive gradient for the oldest OCs group.  
    To extract clearer conclusions about the existence of an azimuthal variation, larger samples are needed.

    \subsection{Dependence with $|Z_{max}|$}

        Several studies have investigated the vertical distribution of abundance with respect to the distance from the Galactic plane ($Z$). A vertical gradient has been reported for field stars \citep[e.g.,][]{Boeche2013,GC_recioblanco2022,Hawkins2023} but not in OCs \citep[e.g.,][]{jacobson2011, Carrera_Pancino2010, carrera2019}. Clusters move along their orbits; hence, here we consider the maximum height of a cluster (|$Z_{\rm max}$|) as a better tracer to perform this study. This orbital parameter has been computed in Paper\,IV. $|Z_{max}|$ is known to be correlated with OC age \citep{Tarricq2021}. This is usually attributed to the vertical heating of the disk, since OCs born in the thin disk, are scattered away from the mid-plane by non-axisymmetric components. 
        This is coupled with the fact that the Galactic potential tends to disrupt OCs, having more chances to survive those that pass more time far away from the disk mid-plane.      

        We do not observe a vertical gradient with [Fe/H], in agreement with previous works. This could be because OCs cover a smaller range of vertical distances (|$Z_{\rm max}$| $\leq$ 1.4 kpc in our sample) compared to field stars. However, we did find positive [X/Fe]-|$Z_{\rm max}$| gradients for Mg, Ti, Al, and Ni, and negative gradients for Na, Sr, Y, Zr, Ba, and Nd. These trends are clearly associated with the dependence of [X/Fe] on age, as discussed in Sect.~\ref{sec:age_gradient}. When we remove the age dependence, the variations with vertical distance disappear.

\section{Summary and conclusions}

In this work, we obtain high-resolution spectra for 194 stars, members of 36 open clusters. We used both equivalent widths and spectral synthesis methods to determine atmospheric parameters and LTE chemical abundances for 21 elements belonging to the main nucleosynthesis groups ($\alpha$, odd-Z, Fe-peak, and neutron-capture elements). We also provide NLTE abundances for elements when corrections are available. Additionally, we construct the OCCASO+ sample by adding the abundances of other 63 clusters studied in similar conditions to ours: high-resolution ($R>$20\,000) and at least four stars sampled in the red-clump region. Both samples are used to investigate abundance trends with $R_{\rm GC}$, and their temporal evolution, azimuth and distance to the plane, from the orbital parameter |Z$_{max}$|, and the abundance dependence with age. The main results of our work are:

\begin{itemize}

\item Some of the studied elements show dependencies with age. [Mg/Fe] and [Si/Fe] show a positive trend towards older ages. Some clusters younger than 0.7 Ga show an unexpected enhancement in those elements. [Na/Fe] shows a decreasing trend until $\sim$ 1.8 Ga and a plateau for older ages, while [Al/Fe] shows a mild positive trend. [Sc/Fe], [Ni/Fe], and [Cu/Fe] show significant positive trends, while  [V/Fe], [Co/Fe] and [Zn/Fe] show negative trends. The neutron capture elements show negative gradients, those of [Zr/Fe] and [Ba/Fe] having a flattening at ages of > 2.5 Ga.

\item We find that the [X/Fe]-age trend of neutron-capture elements depends on their s-process contribution.

\item There is a decreasing [Fe/H] radial gradient and a flattening of the trend beyond 11.3 $\pm$0.8 kpc, the so-called knee shape. Inside the knee radius, the OCCASO and OCCASO+ samples show gradients of -0.06$\pm$0.02 dex kpc$^{-1}$ and -0.07$\pm$0.01 dex kpc$^{-1}$, respectively. Outside the knee, the OCCASO+ sample shows a gradient of -0.03$\pm$0.01 dex kpc$^{-1}$.

\item The radial gradients of the other elements show different tendencies depending on the group to which they belong. $\alpha$ elements have a positive gradient, except [Si/Fe]. Then, [Na/Fe] has a negative gradient, while [Al/Fe] show a flat one. Fe-peak elements have flat gradients, except for [Cr/Fe] and [Zn/Fe] that show positive trends. Neutron capture elements show different behaviors, having [Ba/Fe] and [Nd/Fe] the steepest positive trends.

\item  The [Fe/H] radial gradient shows a steepening with age that seems to be non-linear, since the second age bin (1>age>2 Ga) has a steeper gradient than expected considering the other age bins. We also find that the younger clusters (<1Ga) do not have the knee shape seen in the older ones. This suggests that the event that produced the knee occurred more than 1Ga ago. This, together with the absence of young clusters at more than 14kpc, supports the hypothesis that the knee shape was formed by outward radial migration.
 
\item When examining the temporal evolution of [X/Fe] radial gradient, we generally observe that the trend within each age bin follows the overall trend calculated using all the open clusters. Hence, we do not observe a clear correlation between these trends and the age bin. If there was any evolution of the gradient over time, it would be smaller than our uncertainties. As a result, we can determine an upper limit of $\sim$ 0.01 dex kpc$^{-1}$ for the extent of their change with age. This is, in general, a negligible value of change with age for [X/Fe] gradient. Nevertheless, there are hints that certain elements exhibit an evolution of their radial gradients. Mg appears to show an increase in the gradient with age, while Ti and Ni seem to display a decrease.

\item One feature of the [Fe/H] radial dependence is the large dispersion ($>$0.3\,dex) observed at any position. That is widely attributed to the radial migration, but the observed dispersion could be due to variations of the metal content with Galactic azimuth. We study the existence of the azimuthal gradient by splitting the sample into age bins. We do not find any dependence on azimuth for clusters with ages between 0.1 and 2 Ga. However, we do observe hints of a positive trend for open clusters with ages between 2 and 7.3 Ga.

\item We find $\delta$[X/Fe]/$\delta$|Z$_{max}$| gradients that are mostly due to the trends with age. They disappear once the age dependence is subtracted. 

\end{itemize}

In summary, this work presents high-resolution spectra analysis of 194 stars from 36 open clusters, investigating atmospheric parameters and chemical abundances of 21 elements. The study reveals age-dependent trends in elemental abundances, notably positive trends for [Mg/Fe], [Si/Fe], [Sc/Fe], [Ni/Fe], and [Cu/Fe], and negative trends for [V/Fe], [Co/Fe], [Zn/Fe], [Zr/Fe], and [Ba/Fe]. The radial gradients of [Fe/H] exhibit complex behaviors, with a distinct knee shape observed beyond 11.3 $\pm$0.8 kpc. Additionally, there are hints of azimuthal gradient trends in older clusters and negligible changes in radial gradients with age for most elements.

\begin{acknowledgements}
This work was supported by the MINECO (Spanish Ministry of Economy, Industry and Competitiveness) through grant ESP2016-80079-C2-1-R (MINECO/FEDER, UE) and by the Spanish MICIN/AEI/10.13039/501100011033 and by "ERDF A way of making Europe" by the “European Union” through grants RTI2018-095076-B-C21 and PID2021-122842OB-C21, and the Institute of Cosmos Sciences University of Barcelona (ICCUB, Unidad de Excelencia ’Mar\'{\i}a de Maeztu’) through grant CEX2019-000918-M. 
LC acknowledges the grant RYC2021-033762-I funded by MCIN/AEI/10.13039/501100011033 and by the European Union NextGenerationEU/PRTR.
GT, ES, and AD acknowledge funding from the Lithuanian Science Council (LMTLT, grant No. P-MIP-23-24).
Based on observations made with the Nordic Optical Telescope, owned in collaboration by the University of Turku and Aarhus University, and operated jointly by Aarhus University, the University of Turku and the University of Oslo, representing Denmark, Finland and Norway, the University of Iceland and Stockholm University at the Observatorio del Roque de los Muchachos, La Palma, Spain, of the Instituto de Astrofisica de Canarias.
Based on observations made with the Mercator Telescope, operated on the island of La Palma by the Flemish Community, at the Spanish Observatorio del Roque de los Muchachos of the Instituto de Astrofísica de Canarias. Based on observations obtained with the HERMES spectrograph, which is supported by the Research Foundation - Flanders (FWO), Belgium, the Research Council of KU Leuven, Belgium, the Fonds National de la Recherche Scientifique (F.R.S.-FNRS), Belgium, the Royal Observatory of Belgium, the Observatoire de Genève, Switzerland and the Thüringer Landessternwarte Tautenburg, Germany. Based on observations collected at Centro Astronómico Hispano en Andalucía (CAHA) at Calar Alto, operated jointly by Instituto de Astrofísica de Andalucía (CSIC) and Junta de Andalucía. This research has made use of NASA’s Astrophysics Data System Bibliographic Services.
\end{acknowledgements}


%
%

\bibliographystyle{aa} 
\bibliography{occasov}

\begin{appendix}

\section{Complementary tables and figures}

             \begin{table}[h!]
            \setlength{\tabcolsep}{1mm}
            \begin{center}
            \caption{First rows of the table with the line list for EW. The complete table is available at the CDS.}
            \begin{tabular}{lccccccccc}
            \hline
            Wavelength (nm)&Element\\
            \hline
            480.288&Fe1\\
            480.8148&Fe1\\
            480.9938&Fe1\\
            481.0528&Zn1\\
            481.1983&Ni1\\
            481.4590&Ni1\\
            482.3463&Mn1\\
            482.3483&Mn1\\
            482.3495&Mn1\\
            482.3508&Mn1\\
            \hline
            \end{tabular}
            \label{Tab:linelistEW}
            \end{center}
            \end{table}

            \begin{table}[h!]
            \setlength{\tabcolsep}{1mm}
            \begin{center}
            \caption{First rows of the table with the line list for SS. The complete table is available at the CDS.}
            \begin{tabular}{lccccccccc}
            \hline
            Wavelength (nm)&Element\\
            \hline
            480.1025&Cr1\\
            480.288&Fe1\\
            480.8148&Fe1\\
            481.0528&Zn1\\
            481.1983&Ni1\\
            481.4591&Ni1\\
            482.352&Mn1\\
            483.2426&V1\\
            486.9463&Fe1\\
            487.5493&Ti1\\
            \hline
            \end{tabular}
            \label{Tab:linelistSS}
            \end{center}
            \end{table}

    \begin{table}[h!]
    \setlength{\tabcolsep}{1mm}
    \begin{center}
    \caption{[X/Fe] abundance dependence with age of the samples OCCASO and OCCASO+ and its uncertainties computed in Sect.~\ref{sec:age_gradient}.}
    \begin{tabular}{lccccccc}

    \hline 
    &\multicolumn{3}{c}{OCCASO}&\multicolumn{3}{c}{OCCASO+}\\
    Element&$\delta[X/Fe]/\delta Age$&$N$&$\rho$&$\delta[X/Fe]/\delta Age$&$N$&$\rho$\\
    &[dex Ga$^{-1}$]& & &[dex Ga$^{-1}$]& & \\
    \hline
    Mg&0.009$\pm$0.007&36&0.48&0.012$\pm$0.009&99&0.37\\
    Si&0.012$\pm$0.008&36&0.64&0.011$\pm$0.003&99&0.41\\
    Ca&0.008$\pm$0.005&36&0.15&-0.001$\pm$0.001&99&0.05\\
    Ti&0.009$\pm$0.005&36&0.66&0.005$\pm$0.003&99&0.37\\
    Na&-0.03$\pm$0.013&36&-0.11&-0.005$\pm$0.003&99&-0.12\\
    Al&0.02$\pm$0.011&36&0.61&0.003$\pm$0.002&99&0.49\\
    Sc&-0.005$\pm$0.004&36&-0.43&0.017$\pm$0.005&76&-0.21\\
    V&0.01$\pm$0.005&36&0.62&0.009$\pm$0.004&99&0.18\\
    Cr&0.018$\pm$0.004&36&0.63&0.002$\pm$0.002&99&0.21\\
    Mn&0.003$\pm$0.007&36&0.25&0.005$\pm$0.003&99&0.21\\
    Co&0.007$\pm$0.005&36&0.29&0.01$\pm$0.004&99&0.32\\
    Ni&0.026$\pm$0.006&36&0.77&0.027$\pm$0.005&99&0.54\\
    Cu&0.016$\pm$0.008&36&0.52&0.024$\pm$0.006&78&0.38\\
    Zn&0.02$\pm$0.012&32&0.12&0.01$\pm$0.004&76&0.21\\
    Sr&-0.07$\pm$0.008&35&-0.77&--&--&--\\
    Y&-0.022$\pm$0.012&36&-0.80&-0.011$\pm$0.008&80&-0.35\\
    Zr&-0.027$\pm$0.013&35&-0.7&-0.023$\pm$0.007&78&-0.38\\
    Ba&-0.066$\pm$0.007&36&-0.76&-0.058$\pm$0.011&80&-0.48\\    
    Ce&-0.044$\pm$0.015&36&-0.69&-0.022$\pm$0.008&94&-0.47\\
    Nd&-0.032$\pm$0.011&36&-0.69&-0.016$\pm$0.009&80&-0.11\\  
    
    \hline
    \end{tabular}
            \tablefoot{$N$ and $\rho$ are the number of OCs and the Spearman correlation coefficient, respectively.}
    \label{tab:alpha-abu-age_OCC}
    \end{center}
    \end{table}


            \begin{table*}[h!]
            \setlength{\tabcolsep}{1mm}
            \begin{center}
            \caption{Stellar parameters and chemical abundances for the 194 stars in this work.}
            \begin{tabular}{lccccccccccc}
            \hline
            Cluster&source id {\gaia} DR3&{\teff} (K)&{\logg}&GALAF&$[$Fe/H$]$\textsubscript{EW}&$[$Mg/Fe$]$\textsubscript{SS}&$[$Si/Fe$]$\textsubscript{EW}&$[$Ca/Fe$]$\textsubscript{SS}&$[$Ti/Fe$]$\textsubscript{EW}\\
            \hline
            NGC1907&183263711899696768&5215$\pm$55&2.77$\pm$0.10&1&-0.05$\pm$0.01&0.02$\pm$0.03&0.11$\pm$0.03&0.06$\pm$0.04&0.10$\pm$0.03\\
            NGC1907&183263097725025024&5030$\pm$55&2.60$\pm$0.09&1&-0.03$\pm$0.01&-0.05$\pm$0.03&0.08$\pm$0.03&0.04$\pm$0.03&0.03$\pm$0.03\\
            NGC1907&183263127784145280&5201$\pm$43&2.67$\pm$0.12&1&-0.06$\pm$0.01&0.02$\pm$0.03&0.072$\pm$0.03&0.04$\pm$0.03&0.03$\pm$0.03\\
            NGC1907&183263127784146176&5273$\pm$55&3.04$\pm$0.11&4&-0.03$\pm$0.01&0.03$\pm$0.04&0.06$\pm$0.04&-0.01$\pm$0.04&0.11$\pm$0.03\\
            NGC2099&3451181873619100160&5079$\pm$55&2.73$\pm$0.12&1&0.07$\pm$0.01&-0.07$\pm$0.02&0.13$\pm$0.04&0.01$\pm$0.03&-0.01$\pm$0.03\\
            NGC2099&3451181766240577024&5047$\pm$40&2.60$\pm$0.08&1&0.04$\pm$0.01&-0.07$\pm$0.02&0.08$\pm$0.02&-0.01$\pm$0.03&0.00$\pm$0.03\\
            NGC2099&3451180602308805120&5104$\pm$40&2.78$\pm$0.09&1&0.06$\pm$0.01&-0.07$\pm$0.02&0.05$\pm$0.04&0.02$\pm$0.03&0.01$\pm$0.03\\
            NGC2099&3451181216484770432&4975$\pm$41&2.54$\pm$0.10&1&0.03$\pm$0.01&-0.07$\pm$0.03&0.06$\pm$0.03&-0.01$\pm$0.03&0.00$\pm$0.03\\
            NGC2099&3451181667460701440&5115$\pm$43&2.78$\pm$0.08&1&0.01$\pm$0.01&-0.09$\pm$0.03&0.12$\pm$0.04&0.01$\pm$0.03&0.00$\pm$0.03\\
            NGC2099&3451179949473813376&5052$\pm$46&2.61$\pm$0.06&4&0.06$\pm$0.01&-0.05$\pm$0.03&0.08$\pm$0.03&0.01$\pm$0.03&0.03$\pm$0.03\\
            NGC2099&3451201458669932032&5065$\pm$45&2.68$\pm$0.07&1&0.06$\pm$0.01&-0.07$\pm$0.02&0.09$\pm$0.02&-0.01$\pm$0.03&0.04$\pm$0.03\\

            \hline
                        \vspace{0.5cm}
            \end{tabular}
            \tablefoot{Here, only a sample of rows and columns is shown. The complete table with all the atmospheric parameters and the 21 elements computed by EW and SS methods is  available at the CDS. GALAF is a quality flag assigned to \gala values derived in the study of how the initial guesses of atmospheric parameters affect their derivation (see Sect.~\ref{sec:GalaInit}).}
            \label{tab:AP-ABU}
            \end{center}
            \end{table*}


    \begin{table*}[h!]
    \setlength{\tabcolsep}{1mm}
    \begin{center}
    \caption{Comparison of radial gradients between OCCASO, OCCASO+ and the literature reanalysis of Sect.~\ref{sect:radial_trends}.}
    \begin{tabular}{lccccccccc}
    \hline
    [X/Fe]&Group&OCCASO&OCCASO+&GES23&\citet{myers2022}&\citet{spina2021}\\
    &&$\delta$[X/Fe]/$\delta R_{\rm GC}$&$\delta$[X/Fe]/$\delta R_{\rm GC}$&$\delta$[X/Fe]/$\delta R_{\rm GC}$&$\delta$[X/Fe]/$\delta R_{\rm GC}$&$\delta$[X/Fe]/$\delta R_{\rm GC}$\\
    &&[dex kpc$^{-1}$]&[dex kpc$^{-1}$]&[dex kpc$^{-1}$]&[dex kpc$^{-1}$]&[dex kpc$^{-1}$]\\
    \hline
    Mg&$\alpha$&0.017$\pm$0.011&0.006$\pm$0.005&0.008$\pm$0.006&0.001$\pm$0.001&0.001$\pm$0.002\\
    Si&$\alpha$&0.008$\pm$0.006&0.000$\pm$0.003&0.000$\pm$0.001&-0.0$\pm$0.001&0.0$\pm$0.001\\
    Ca&$\alpha$&0.013$\pm$0.005&0.016$\pm$0.004&0.033$\pm$0.006&0.009$\pm$0.002&-0.003$\pm$0.003\\
    Ti&$\alpha$&0.016$\pm$0.007&0.025$\pm$0.006&0.024$\pm$0.006&0.009$\pm$0.006&0.011$\pm$0.008\\
    \hline
    Na&Odd-Z&-0.027$\pm$0.008&-0.021$\pm$0.006&-0.018$\pm$0.007&-0.058$\pm$0.014&-0.007$\pm$0.004\\
    Al&Odd-Z&-0.013$\pm$0.007&-0.001$\pm$0.005&0.017$\pm$0.007&0.005$\pm$0.004&-0.002$\pm$0.001\\
    \hline
    Sc&Fe-peak&0.001$\pm$0.007&0.01$\pm$0.003&0.007$\pm$0.005&&\\
    V&Fe-peak&-0.01$\pm$0.006&0.005$\pm$0.003&0.013$\pm$0.005&-0.0$\pm$0.001&\\
    Cr&Fe-peak&0.017$\pm$0.008&0.02$\pm$0.006&0.020$\pm$0.005&-0.001$\pm$0.005&\\
    Mn&Fe-peak&-0.007$\pm$0.010&0.002$\pm$0.003&0.006$\pm$0.005&-0.017$\pm$0.004&-0.008$\pm$0.002\\
    Co&Fe-peak&-0.007$\pm$0.005&-0.001$\pm$0.002&0.003$\pm$0.007&-0.039$\pm$0.012&-0.014$\pm$0.004\\
    Ni&Fe-peak&0.000$\pm$0.004&-0.003$\pm$0.002&-0.001$\pm$0.005&-0.001$\pm$0.003&-0.01$\pm$0.003\\
    Cu&Fe-peak&-0.005$\pm$0.010&0.009$\pm$0.008&0.041$\pm$0.009&&\\
    Zn&Fe-peak&0.026$\pm$0.012&0.015$\pm$0.004&0.001$\pm$0.004&&\\
    \hline
    Sr&n-capture&0.007$\pm$0.012&&&&\\
    Y&n-capture&-0.005$\pm$0.011&0.015$\pm$0.009&0.003$\pm$0.009&&\\
    Zr&n-capture&0.009$\pm$0.015&0.015$\pm$0.011&0.015$\pm$0.007&&\\
    Ba&n-capture&0.026$\pm$0.018&0.032$\pm$0.014&0.025$\pm$0.007&&\\
    Ce&n-capture&0.001$\pm$0.013&0.017$\pm$0.007&0.007$\pm$0.008&0.065$\pm$0.013&\\
    Nd&n-capture&0.032$\pm$0.016&0.05$\pm$0.008&0.061$\pm$0.015&&\\
    \hline
    
    \end{tabular}
    \label{tab:comp}
    \end{center}
    \end{table*}

    \begin{table*}[h!]
    \setlength{\tabcolsep}{1mm}
    \begin{center}
    \caption{[X/Fe] radial gradients studied through MCMC at Sect.~\ref{sect:radial_trends} in the OCCASO+ sample. }
    \begin{tabular}{lccccccccc}
    \hline
    Element&b1&m1&m2&Knee&N&Spearman\\   
     &[dex]&[dex kpc$^{-1}$]& [dex kpc$^{-1}$]&[kpc]& \\
    \hline
    Mg&-0.118$\pm$0.021&0.008$\pm$0.002&&&99&0.29\\
    Si&0.029$\pm$0.018&0.005$\pm$0.002&&&99&0.21\\
    Ca&-0.07$\pm$0.022&0.008$\pm$0.002&&&99&0.42\\
    Ti&-0.145$\pm$0.047&0.019$\pm$0.006&-0.003$\pm$0.009&10.7$\pm$1.1&99&0.40\\
    Na&0.287$\pm$0.082&-0.026$\pm$0.01&0.015$\pm$0.008&9.9$\pm$0.8&99&-0.28\\
    Al&-0.034$\pm$0.029&0.005$\pm$0.003&&&99&0.19\\
    Sc&0.001$\pm$0.019&0.008$\pm$0.002&&&76&0.38\\
    V&0.02$\pm$0.022&-0.002$\pm$0.002&&&99&0.09\\
    Cr&-0.246$\pm$0.058&0.029$\pm$0.007&-0.008$\pm$0.005&9.5$\pm$0.5&99&0.33\\
    Mn&-0.146$\pm$0.019&0.003$\pm$0.002&&&99&0.04\\
    Co&0.019$\pm$0.027&0.002$\pm$0.003&&&99&0.11\\
    Ni&0.008$\pm$0.018&-0.003$\pm$0.002&&&99&-0.11\\
    Cu&-0.235$\pm$0.054&0.003$\pm$0.006&&&78&0.13\\
    Zn&-0.152$\pm$0.021&0.01$\pm$0.002&&&76&0.54\\
    Y&-0.129$\pm$0.037&0.011$\pm$0.004&&&80&0.38\\
    Zr&-0.353$\pm$0.079&0.038$\pm$0.01&-0.02$\pm$0.007&9.6$\pm$0.4&78&0.18\\
    Ba&-0.321$\pm$0.111&0.036$\pm$0.014&-0.016$\pm$0.018&10.3$\pm$1.2&80&0.25\\
    Ce&-0.107$\pm$0.075&0.027$\pm$0.009&0.006$\pm$0.012&10.7$\pm$0.9&94&0.36\\
    Nd&-0.382$\pm$0.074&0.049$\pm$0.009&0.016$\pm$0.016&10.7$\pm$1.3&80&0.71\\
    \hline
    
    \end{tabular}
            \tablefoot{The columns b1 and m1 are the y-intercept and slope of the first line fitted, respectively. The column m2 is the slope of the second line. The position of the knee, the number of OCs and Spearman correlation coefficient are shown in the last columns.}
    \label{tab:OPLUS_all}
    \end{center}
    \end{table*}

    \begin{table*}[h!]
    \setlength{\tabcolsep}{1mm}
    \begin{center}
    \caption{Change with age of the [Fe/H] radial and azimuthal gradients studied with a multilinear regression at Sect.~\ref{sect:az}.}
    \begin{tabular}{lcccccc}
    \hline 

 & &Radial gradient&\multicolumn{2}{c}{Radial and azimuthal gradients} & & \\

Sample&Age bin&$\delta$[Fe/H]/$\delta R_{\rm GC}$&$\delta$[Fe/H]/$\delta\Phi$&$N$&Spearman\\

&[Ga]& [dex kpc$^{-1}$]& [dex deg$^{-1}$]& & \\

    \hline
OCCASO&0.1 - 1& -0.022$\pm$0.007&-0.0034$\pm$0.0016&15&0.33\\
OCCASO&1 - 2& -0.069$\pm$0.016&0.0048$\pm$0.0026&11&0.83\\
OCCASO&2 - 3& -0.026$\pm$0.038&-0.0024$\pm$0.0049&4&0.34\\
OCCASO&3 - 7.3&  -0.101$\pm$0.019&0.0003$\pm$0.0043&5&0.87\\
    \hline
OCCASO+&0.1 - 1&-0.051$\pm$0.004&0.0001$\pm$0.0007&39&0.79\\
OCCASO+&1 - 2&-0.060$\pm$0.004&-0.0003$\pm$0.0007&29&0.87\\
OCCASO+&2 - 3&-0.041$\pm$0.004&0.0026$\pm$0.0008&14&0.92\\
OCCASO+&3 - 7.3&-0.088$\pm$0.008&0.003$\pm$0.0019&15&0.83\\
    \hline
\citet{myers2022} &0.1 - 1&-0.053$\pm$0.002&0.0016$\pm$0.0002&39&0.74\\
\citet{myers2022} &1 - 2&-0.056$\pm$0.002&0.0000$\pm$0.0004&20&0.91\\
\citet{myers2022} &2 - 3&-0.026$\pm$0.005&0.0063$\pm$0.0007&11&0.92\\
\citet{myers2022} &3 - 7.3&-0.046$\pm$0.002&0.0074$\pm$0.0006&12&0.84\\
    \hline
\citet{spina2021}&0.1 - 1&-0.064$\pm$0.002&0.0007$\pm$0.0002&57&0.80\\
\citet{spina2021}&1 - 2&-0.062$\pm$0.005&0.0003$\pm$0.0007&22&0.77\\
\citet{spina2021}&2 - 3&-0.041$\pm$0.003&0.0052$\pm$0.0005&10&0.97\\
\citet{spina2021}&3 - 7.3&-0.060$\pm$0.003&0.0037$\pm$0.0009&16&0.78\\
    \hline
GES23 &0.1 - 1&-0.044$\pm$0.006&-0.0014$\pm$0.0011&23&0.76\\
GES23 &1 - 2&-0.058$\pm$0.005&0.0005$\pm$0.0011&17&0.88\\
GES23 &2 - 3&-0.048$\pm$0.015&0.0057$\pm$0.0026&7&0.86\\
GES23 &3 - 7.3&-0.038$\pm$0.004&0.0072$\pm$0.0012&12&0.77\\
    \hline
\gaia DR3&0.1 - 1&-0.050$\pm$0.003&0.0002$\pm$0.0004&303&0.43\\
\gaia DR3&1 - 2&-0.056$\pm$0.008&-0.0012$\pm$0.0008&48&0.52\\
\gaia DR3&2 - 3&-0.079$\pm$0.012&0.0000$\pm$0.0014&25&0.57\\
\gaia DR3&3 - 7.3&-0.067$\pm$0.019&0.0041$\pm$0.0017&13&0.71\\

    \hline
    \end{tabular}
            \tablefoot{We analyzed the OCCASO and OCCASO+ samples and reanalyzed the others. Number of OCs and Spearman correlation coefficient (sixth and seventh columns).}
    \label{tab:AZT}
    \end{center}
    \end{table*}


             \label{fig:separated_age_bins}

         \begin{figure*}
          \includegraphics[width=18.5cm]{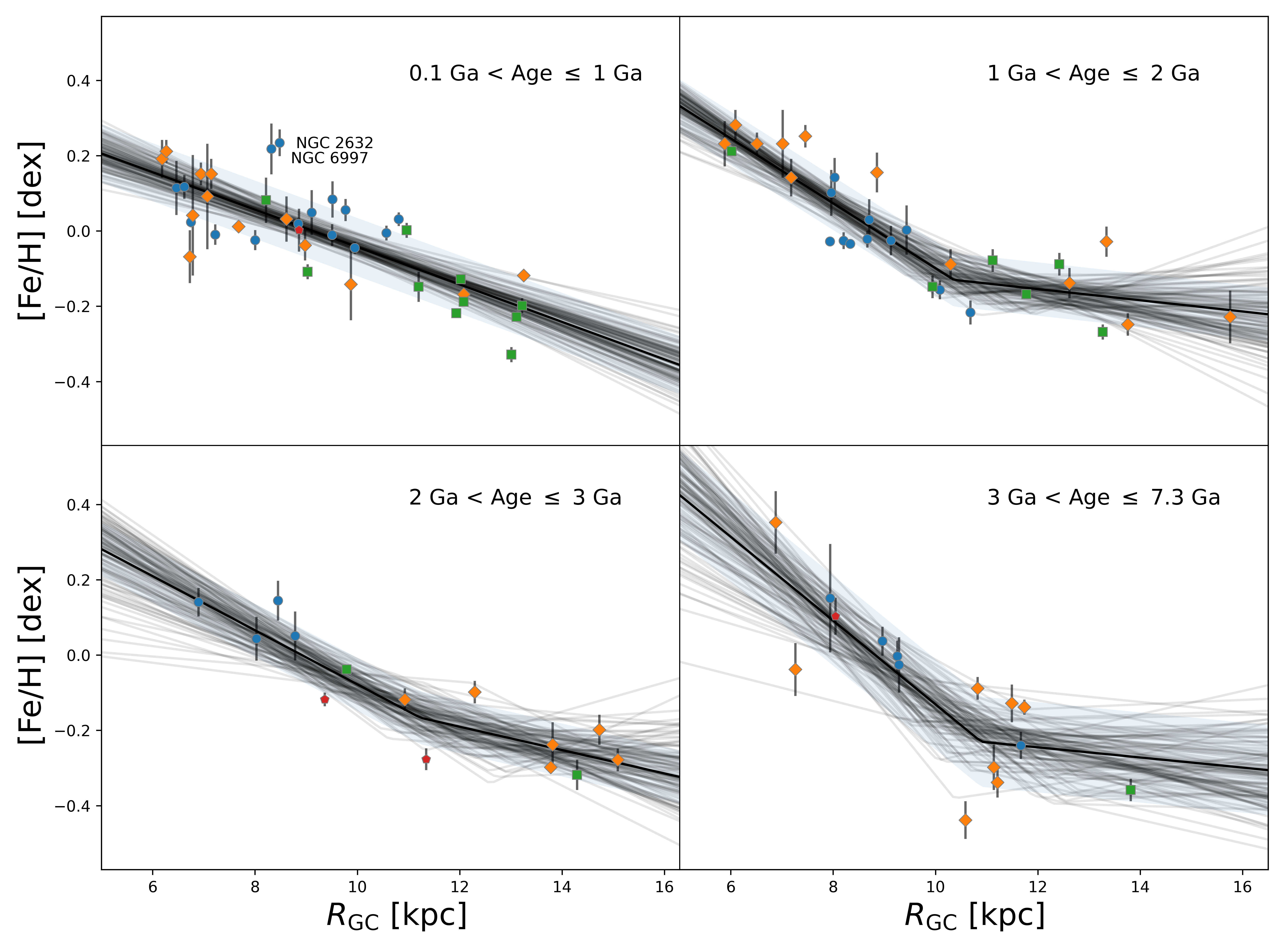}
             \caption{[Fe/H] versus Galactocentric radius for OCCASO+ sample separated in four age bins 0.1-1, 1-2, 2-3, >3 Ga. The different surveys are color-coded as in Fig.~\ref{fig:XY}. The grey vertical lines represent the uncertainties, and the black line is our best fit.}           
             \label{fig:separated_age_bins}
        \end{figure*}

         \begin{figure*}

          \includegraphics[width=18.5cm]{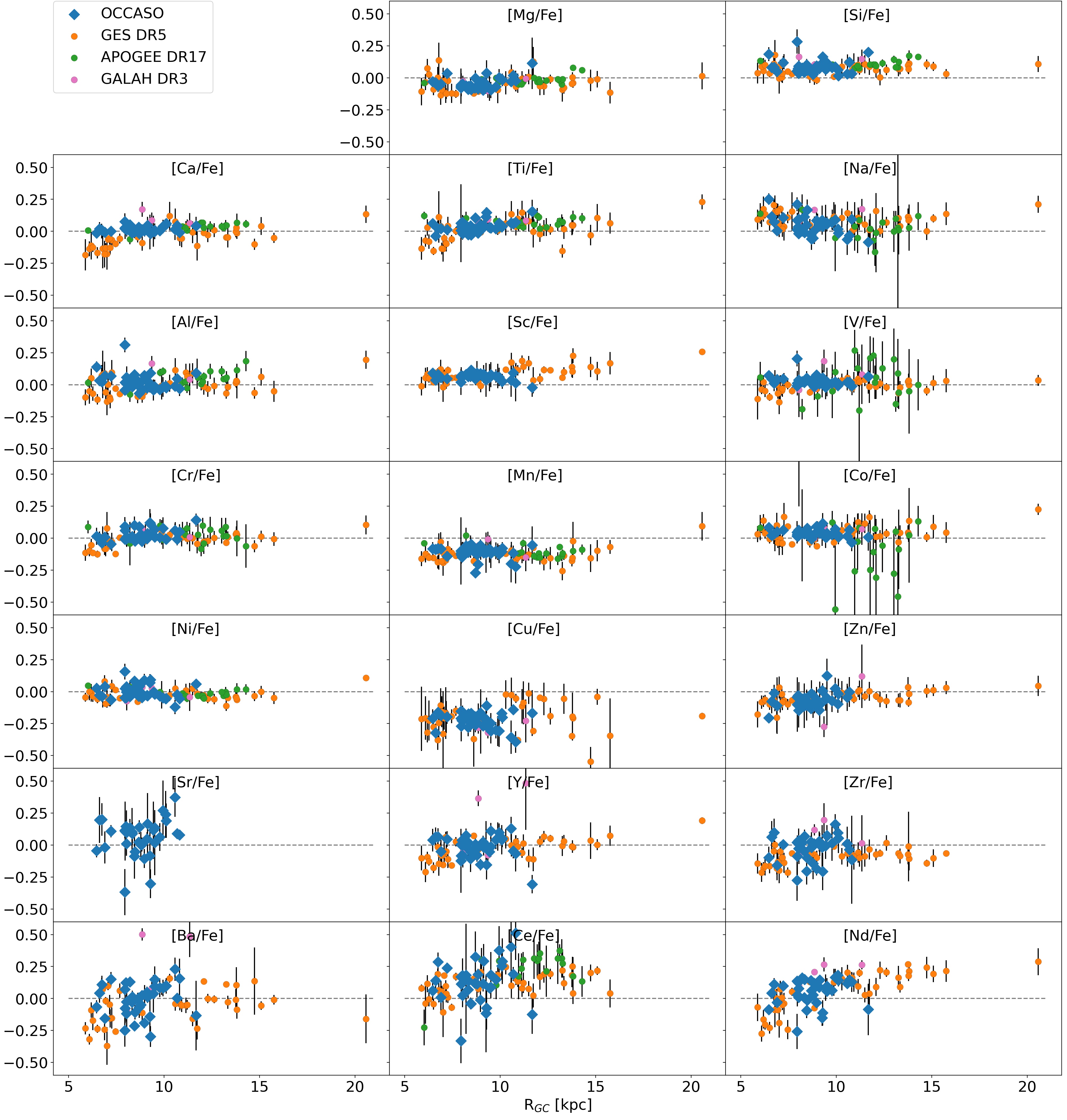}

             \caption{Dependence of [X/Fe] on Galactocentric radius for the clusters in the OCCASO+ sample. The original sources of the abundances are indicated with different symbols and colors.}
             \label{fig:AnexSComp}
        \end{figure*}
        
        \begin{figure*}
        \centering
        \includegraphics[width=10.7cm]{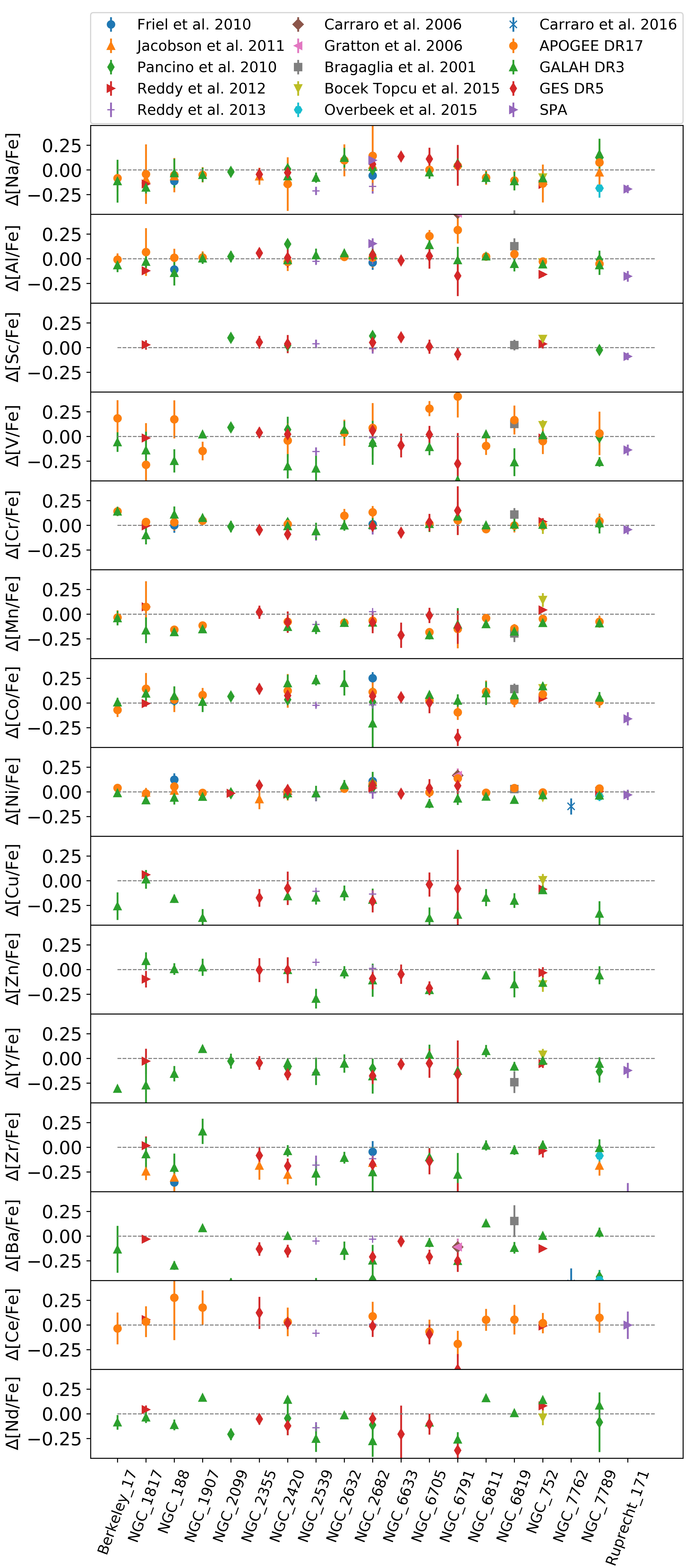}
        \caption{Comparison of the abundances for OCs in common with high-resolution ($R > $20\,000) spectroscopic studies (this work-literature) for elements not shown in Fig. \ref{fig:litOC}.  }
        \label{fig:litOC_annex}
        \end{figure*}

\end{appendix}

\end{document}